\documentclass[%
twocolumn,
 amsmath,amssymb,
aps,
prx,
]{revtex4-1}

\usepackage{graphicx,epsfig,epstopdf}
\usepackage{amsmath}
\usepackage{color}
\usepackage{caption}
\usepackage{subcaption}

\usepackage{tabularx}

\def\e{\begin{equation}}
\def\f{\end{equation}}
\def\_#1{{\bf #1}}
\def\.{\cdot}

\begin{document}

\title{\textcolor{black}{Asymmetric Metal-dielectric Meta-cylinders and their Potential Applications from  \textcolor{black}{Engineering Scattering Patterns} to Spatial Optical Signal Processing} } 

\author{Ali Momeni$^1$, Mahdi Safari$^2$, Ali Abdolali$^1$, Nazir P. Kherani$^{2,3}$, Romain Fleury$^4$}
 \affiliation{%
 	$^1$Applied Electromagnetic Laboratory, School of Electrical Engineering, Iran University of Science and Technology, Narmak, Tehran, Iran\\
 	$^2$Department of Electrical and Computer Engineering, University of Toronto, 10 King’s College Road, Toronto, M5S 3G4, Canada\\
 	$^3$Department of Materials Science and Engineering, University of Toronto, 140 College Street, 
 	Toronto, ON M5S 3E4, Canada\\
 	$^4$Laboratory of Wave Engineering,
Swiss Federal Institute of Technology in Lausanne (EPFL), CH-1015 Lausanne, Switzerland
 }
\begin{abstract}
We propose a novel type of bi-anisotropic hybrid metal-dielectric structure comprising dielectric and metallic cylindrical wedges wherein the composite meta-cylinder enables advanced control of electric, magnetic and magnetoelectric resonances. We establish a theoretical framework in which the electromagnetic response of this meta-atom is described through the electric and magnetic \textcolor{black}{ multipole moments}. The complete dynamic polarizability tensor, expressed in a compact form, is derived as a function of the Mie scattering coefficients. Further, the constitutive parameters -determined analytically- illustrate the tunability of the structure's frequency and strength of resonances in light of
its high degree of geometric freedom. \textcolor{black}{ Flexibility in the design makes the proposed meta-cylinder a viable candidate for various applications in the  microscopic (single meta-atom) and macroscopic (metasurface) levels. We show that the highly versatile bi-anisotropic meta-atom is amenable to being designed for the desired electromagnetic response, such as electric dipole-free and zero/near-zero (backward and forward) scattering at the microscopic level. In addition, we show that the azimuthal asymmetry gives rise to normal polarizability components which are vital elements in synthesizing asymmetric Optical Transfer Function (OTF) at the macroscopic level.} \textcolor{black}{We conduct a precise inspection, from the microscopic to the macroscopic level, of the metasurface synthesis for emphasizing on the role of normal polarizability components for spatial optical signal processing. It is shown that this simple two-dimensional asymmetric meta-atom can perform  first-order differentiation and edge detection at normal illumination.}  The results reported herein contribute toward improving the physical understanding of wave interaction with artificial materials composed of asymmetric elongated metal-dielectric inclusions and open the potential of its application in spatial signal and image processing.\end{abstract}

\maketitle

\section{Introduction}
~~~Modeling of two-dimensional structures illuminated by electromagnetic waves has been of leading research interest for a long time. The problem of deriving the polarization of elongated structures has been of central interest in the study of antenna theory over the last century  \cite{balanis}. Due to their simplicity and practicality, long metallic and dielectric cylinders and wire media structures have received much attention among all the various canonical 2D structures. These include structures realizing artificial plasmas  \cite{rotman}, hyperbolic
media \cite{Smith,Lu,Poddubny,Simovski}, creating exotic material properties  \cite{Morgado,Pendry,Silveirinha,Laroche,Ghenuche,McMahon}, including those for antenna applications \cite{Ikonen,Capolino}, tailoring the phase of reflection waves in metasurface applications  \cite{Salary,momeni,Zhang,Pors,Lin,Yanik,Rajabalipanah}, imaging and endoscopy  \cite{Casse,Shvets,Silveirinha1}, manipulating Casimir forces \cite{Maslovski}, enhancing the coupling to quantum sources  \cite{Yao,Jacob,Cortes} and enabling single-molecule biosensors \cite{Kabashin}.

The first fundamental step to describing the EM response of a metamaterial is to model the response of an individual particle. In this method each inclusion serves as a polarizable particle which is modeled with a pair of electromagnetic polarizable dipoles, which in turn becomes a new source of electromagnetic fields that lead to corresponding local fields. Ultimately, these effects form the macroscopic constitutive parameters. Unusual properties of this material are observed near resonance and its dependence on the geometry and EM properties of the individual inclusions are self-evident  \cite{Kallos}.
\begin{figure*}[t!]
	\centering\includegraphics[width=18cm]{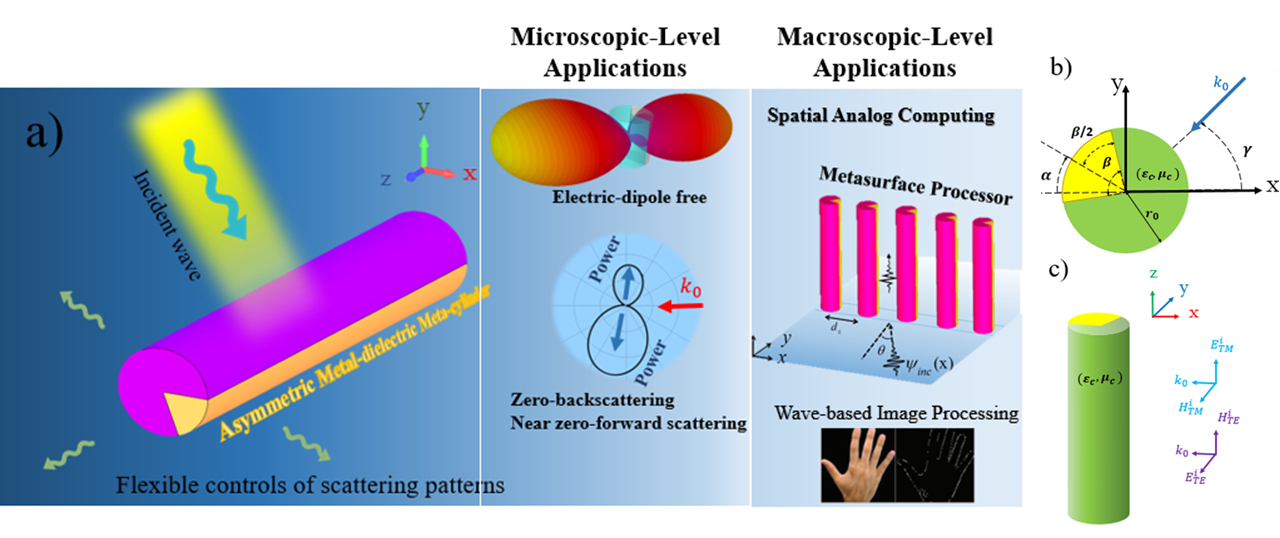}
	\caption{a) The schematic sketch of the proposed bianisotropic
		meta-atom and its microscopic and macroscopic-level applications. b) 2D view, and c) 3D view of meta-atom orientation.}
\end{figure*}
A number of recent studies retrieve the EM response for elongated cylinders under specific conditions of incidence, such as normal incidence \cite{Kallos}. For example, static expressions for the transverse polarizability components of circular cylinders are well established, and recently have been extended to the dynamic case based on far-field scattering considerations  \cite{Strickland}. 
However, because of symmetry restrictions of the infinitely long cylinder, this particle does not possess any magnetoelectric coupling response at normal incidences. In general, no magnetoelectric coupling may exist in the microscopic polarizability of a scatterer with both temporal and inversion symmetry  \cite{Raab,Barron}. Most structures proposed so far in this area act only as a symmetric system which gives almost no flexibility for controlling the scattering patterns. Considering the inability of these symmetric meta-atoms, exploring an asymmetric meta-atom with different geometrical parameters is of high  relevance.

Moreover, computational metasurfaces have been explored to efficiently manipulate the optical wave front for spatial optical analog computing \cite{silva2014performing}. The advantage of metasurfaces over the conventional electronic systems for analog computing and signal processing is two-fold, lower energy consumption and lower computational time. The recognition of these advantages makes us believe that single processing with metasurfaces is a promising approach toward the realization of ultrafast computing systems \cite{pors2015analog}. Over the past few years, several metamaterial/metasurfce-based computational devices have been reported in optical analog computing realm \cite{babaee2020parallel,kwon2018nonlocal,zhu2017plasmonic,momeni2,momeni3,cordaro2019high,he2020spatial,zhu2019generalized,kwon2020dual,lou2020optical}. Metasurface-based processors have been gaining significant attention so as to mitigate the drawbacks associated with the Fourier transform sub-blocks, such as topological analog signal processing \cite{zangeneh2019topological}, photonic crystals \cite{guo2018photonic}, and all-dielectric metasurfaces \cite{zhou2020analog,cordaro2019high}. 

Here, we introduce the metal-dielectric 'meta-cylinder' as an asymmetric meta-atom with a tunable wedge angle presents high degrees of design freedom. We firstly analyze the metal-dielectric meta-cylinder with large degrees of freedom, and thereafter we derive the full dynamic polarizability tensors describing the electromagnetic response of the proposed meta-atom. Breaking azimuthal symmetry can result in an unconventional scattering pattern and unbalanced radiation power which also can be useful for tailoring the transverse optical forces. The fact that asymmetrical structures arise excitation of normal polarizability tensor components (i.e., towards the direction of illumination), opens up new possibilities for pursuing different applications such as beam deflection and advanced analog computing \cite{momeni2,momeni3,kivshar}, where designing and realizing a metasurface processor with  an asymmetrical Optical Transfer Function (OTF) capable of distinguish and separating the $k_t$ and $-k_t$ at normal incident is one of the main challenges of this field \cite{kwon2018nonlocal,zhu2019generalized,davis2019metasurfaces,momeni2,momeni3}. To demonstrate the ability of the asymmetrical metal-dielectric meta-cylinder as an adjustable 2D meta-atom, we utilize the proposed meta-atom to address the recent challenges of the field of wave engineering and signal processing. The optical performance of the proposed meta-atom at both  single meta-atom and metasurface levels is explored, in which near-zero scattering achieved from a single meta-atom showing its viability for invisible sensors and transparent metamaterial applications and a metasurface comprising an array of meta-cylinder is proposed for signal and image processing. A metasurface processor consisting of the proposed meta-atom is realised which is well capable of performing mathematical operations such as first-order differentiation operation, edge detection of an arbitrary input signal.

\

\section{SCATTERING ANALYSIS OF A METAL-DIELECTRIC META-CYLINDER }
~~~The configuration of the structure and illumination is shown in  Fig. 1. The presented meta-atom consists of two finite wedges which together form a complete cylinder. The material properties for these wedges are different, one is a metal and the other is a dielectric where $\varepsilon_{c}$  and ${{\mu }_{c}}$ are its permittivity and permeability, respectively.

${{r}_{0}}$ is the radii of the structure and $\beta$ is the metal wedge angle. In order to solve the 2D problem in the most general manner, we consider the incident wave to impinge upon the structure at different azimuthal angles. This is achieved by fixing the incident wave while rotating the structure to the desired angle ($\alpha$)  (see \textcolor{blue}{Fig. 1}).  It is noteworthy that in this study we assume the time dependence to be $e^{-j\omega t}$.\

Considering that the structure is of infinite extent along the longitudinal axis it is clear that \textcolor{black}{ $\frac{\partial F}{\partial z}=0$ where F represent all the EM parameters.} Given that the problem of EM scattering from a finite PEC wedge has been solved previously for TE and TM polarizations \cite{Klinkenbusch}, here we extend the structure by filling-in the remaining part of the cylinder with a dielectric and thus enabling the analysis to simply avail the same scattering wave forms.  \\ 
\begin{align}
&E_{z,sc}^{TM}=\\
&\sum\limits_{n=0}^{\infty }{\bigg(a_{n}^{TM}\cos (}n(\varphi +\alpha ))+b_{n}^{TM}\sin (n(\varphi +\alpha ))\bigg)H_{n}^{(1)}({{k}_{0}}r)\nonumber
\end{align}
\begin{align}
&H_{z,sc}^{TE}=\\
&\sum\limits_{n=0}^{\infty }{\bigg(a_{n}^{TE}\cos (}n(\varphi +\alpha ))+b_{n}^{TE}\sin (n(\varphi +\alpha ))\bigg)H_{n}^{(1)}({{k}_{0}}r)\nonumber
\end{align}
where $E^{TM}_{z,sc}$ and $H^{TE}_{z,sc}$ are the electric and magnetic field components along the $z$ axis describing the $TM$ and $TE$ polarizations, respectively. $k_0=\omega \sqrt{\epsilon_0 \mu_0}$ denotes the wavenumber in free space and $\alpha$ is the orientation angle of meta-cylinder. It is worth noting that the preceding equations are derived using the EM boundary conditions, in which the $a_{n}$ and $b_{n}$ are the Mie scattering coefficients. A detailed derivation of the aforementioned equations is outlined in the \textbf{Appendix A and B}. Assuming that the radius of the meta-atom is smaller than the wavelength, the higher order modes can be neglected leaving the dominant first order terms. Hence, replacing the Hankle functions with its far-field approximation simplifies the scattering wave expression as shown below.
\begin{equation}
\begin{aligned}
&E_{z}^{\text{TM,far}}=\\
&\sqrt{\frac{2}{\pi {{k}_{0}}r}} \bigg\{a_{0}^{TM}+\bigg(a_{1}^{TM}\cos (\alpha )+b_{1}^{TM}\sin (\alpha )\bigg) \cos (\varphi )+\\
&\bigg(b_{1}^{TM}\cos (\alpha )-a_{1}^{TM}\sin (\alpha )\bigg)\sin (\varphi )\bigg\}\exp \bigg(j({{k}_{0}}r-\pi /4)\bigg) 
\end{aligned}
\end{equation}
\begin{equation}
\begin{aligned}
&H_{z}^{\text{TE,far}}=\\
&\sqrt{\frac{2}{\pi {{k}_{0}}r}} \bigg\{a_{0}^{TE}+ \bigg(a_{1}^{TE}\cos (\alpha )+b_{1}^{TE}\sin(\alpha )\bigg)\cos(\varphi )+  \\
&\bigg(b_{1}^{TE}\cos (\alpha )-a_{1}^{TE}\sin (\alpha )\bigg)\sin (\varphi )\bigg\}\exp \bigg(j({{k}_{0}}r-\pi /4)\bigg)
\end{aligned}
\end{equation}
\begin{figure}[t!]
	\centering
	\centering\includegraphics[width=9cm]{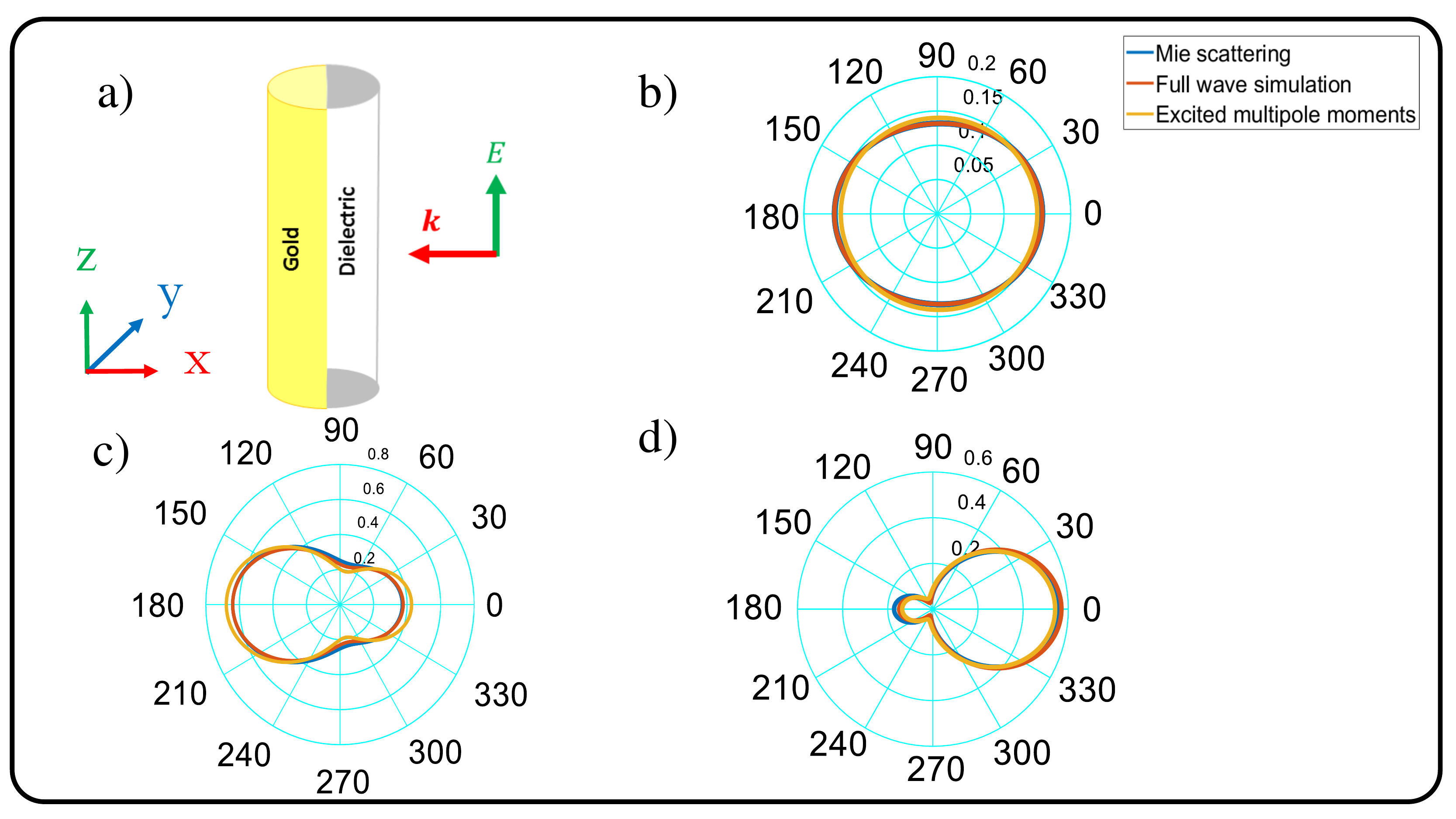}
	\caption{ a) Schematic showing an metal-dielectric meta-cylinder ($\epsilon_r=18$) with wedge angle of $\pi$ which is excited by TM-polarized waves propagating on the -x 
		direction. Far-field scattering pattern from a mentioned meta-atom with different radius b) 600 nm, c) 800 nm, and d) $1\mu m$ using Mie scattering, full wave simulation and excited multipole moments.   }
\end{figure}
For the thin metal-dielectric meta-cylinder which are of interest here, scattering is dominated by the n = 0 and 1 in TM and  TE harmonics, that obviously depending on the incident polarization, the geometric ($\alpha $ and $\beta $ angles and ${{r}_{0}}$) and constitutive parameters (${{\varepsilon }_{c}={\varepsilon }_{r}{\varepsilon }_{0}}$ and ${{\mu }_{c}={\mu }_{r}{\mu }_{0}}$ of dielectric wedge).
\section{DYNAMIC POLARIZABILITY TENSOR EXTRACTION }
Modelling of the electromagnetic fields using multipole moments has several advantages. The electromagnetic fields are linearly dependent on the moments and as a result do not involve complicated integrals \cite{Kallos}. The resulting multiple moments are vital elements in the description of homogeneous effective medium theories such as Maxwell-Garnet and other nonlocal methods \cite{Zouhdi,Silveirinha1}. In the following sections, we first derive the analytical expression for the induced electric and magnetic dipoles based on the multiple moments theory. Next, we extract the polarizability tensors using the standing wave approach which describe the EM interaction of the proposed structure. 
\subsection{Extraction of multipole moments}
The radiated fields due to a set of electric and magnetic dipoles can be approximately expressed as presented by \cite{Papas}
\begin{align}
& \vec{E}\simeq {{Z }_{0}}\bigg\{[k_{0}^{2}\vec{p}{{c}_{0}}G+(\vec{p}{{c}_{0}}.\nabla )\nabla G]+j{{k}_{0}}[\nabla G\times \vec{m}]\bigg\}, \\
& \vec{H}\simeq -j{{k}_{0}}[\nabla G\times \vec{p}{{c}_{0}}]+[k_{0}^{2}\vec{m}G+(\vec{m}.\nabla )\nabla G],
\end{align}
where ${{k}_{0}}$, ${{c}_{0}}$ are the wave vector and the speed of light, respectively. $\nabla =\hat{r}\frac{\partial }{\partial r}$ ($\hat{r}$ is a cylindrical unit vector along the radial direction).
Here, we are going to derive the induced electric and magnetic dipole moments in a 2D structure. Consequently, we can use the 2D cylindrical Green function  $G=\frac{j}{4}H_{0}^{(1)}\left( {{k}_{0}}r \right)$ as an alternative. As a result, we utilize the analytical expression for the electromagnetic radiation of electric and magnetic dipole moments wherein the radiated fields are expressed as TE and TM modes \cite{Kallos} using far field approximation.
\begin{align}
& \vec{E}_\text{TM}^\text{far}=\hat{z}\frac{j{{Z }_{0}}k_{0}^{2}}{4}\sqrt{\frac{2}{\pi {{k}_{0}}r}}\left( {{p}_{z}}{{c}_{0}}-2{{m}_{\varphi }} \right)\exp \bigg(j({{k}_{0}}r-\pi /4)\bigg),\\
& \vec{H}_\text{TM}^\text{far}=-\frac{jk_{0}^{2}}{4}\hat{\varphi }\left( {{p}_{z}}{{c}_{0}}-2{{m}_{\varphi }} \right)\exp \bigg(j({{k}_{0}}r-\pi /4)\bigg),\\
& \vec{E}_\text{TE}^\text{far}=\hat{\varphi }\frac{j{{Z }_{0}}k_{0}^{2}}{4}\sqrt{\frac{2}{\pi {{k}_{0}}r}}\left( {{p}_{\varphi }}{{c}_{0}}+{{m}_{z}} \right)\exp \bigg(j({{k}_{0}}r-\pi /4)\bigg),\\
& \vec{H}_\text{TE}^\text{far}=\frac{jk_{0}^{2}}{4}\hat{z}\left( {{p}_{\varphi }}{{c}_{0}}+{{m}_{z}} \right)\exp \bigg(j({{k}_{0}}r-\pi /4)\bigg),
\end{align}
~~~It is noteworthy,  that $p_{\phi}$ and $m_{\phi}$ describe projections of the moments along the azimuthal direction ($p_\phi = p_y cos\phi - p_x sin\phi$, $m_{\phi} = m_y cos \phi - m_x sin\phi$) and should not be confused with the azimuthal components since the dipoles \textbf{\textit{P}} and \textbf{\textit{M}} are located at the origin of the axes. 
Observing the above equations, it is clear that the radiating fields correspond to typical cylindrical TEM waves, satisfying $\vec{H}=\hat{n}\times E/{{Z }_{0}}$.   
The dipole moments excited in the metal-dielectric meta-cylinder can now be derived using the approach presented in \cite{Kallos}. First, we analytically extract the far-field scattering TE and TM fields for the metal-dielectric meta-cylinder (see section II), and then we derive the far-field radiation due to the set of electric and magnetic dipoles (\textcolor{blue}{Eqs. 12} to \textcolor{blue}{15}). Thereafter, we extract the induced dipole moments by comparing the \textcolor{blue}{Eqs. 3} and \textcolor{blue}{4} and \textcolor{blue}{Eqs. 12} to \textcolor{blue}{15}.\

The relation between the Mie coefficients and the induced dipole moments is derived:
\begin{align} P=&-\hat{x}\frac{4\bigg( b_{1}^{TE}\cos (\alpha )-a_{1}^{TE}\sin (\alpha )\bigg)}{jk_{0}^{2}{{c}_{0}}}\\
&+\hat{y}\frac{4\bigg( a_{1}^{TE}\cos (\alpha )+b_{1}^{TE}\sin (\alpha ) \bigg)}{jk_{0}^{2}{{c}_{0}}}+\hat{z}\frac{4a_{0}^{TM}}{jk_{0}^{2}{{Z}_{0}}{{c}_{0}}} \\\nonumber
M=&\hat{x}\frac{2\bigg( b_{1}^{TM}\cos (\alpha )-a_{1}^{TM}\sin (\alpha ) \bigg)}{jk_{0}^{2}{{Z}_{0}}}\\
&-\hat{y}\frac{2\bigg( a_{1}^{TM}\cos (\alpha )+b_{1}^{TM}\sin (\alpha ) \bigg)}{jk_{0}^{2}{{Z}_{0}}}+\hat{z}\frac{4a_{0}^{TE}}{jk_{0}^{2}} \nonumber\end{align}
~~~Upon observation of the equations, it is clear that the extracted dipole moments can be easily tuned by varying the angular rotation $\alpha$ and the scattering Mie-coefficients by changing  the wedge angle $\beta$.
In order to verify the validity of our approach, the far-field scattering pattern of a metal-dielectric meta-cylinder ($\epsilon_r=18$) with wedge angle of $\pi$ was derived using the full-wave simulation (CST Microwave Studio), the radiation of the induced dipole moments, and also the closed-form Mie scattering of the metal-dielectric meta-cylinder (section II). The incident wave is assumed to be a TM plane wave  with the operating wave length of ${{\lambda }_{0}}=6\mu m$. \textcolor{blue}{Figure. 2} presents the mentioned results for three metal-dielectric cylinders with different radii’s 600 $nm$, 800 $nm$, and 1$\mu m$, respectively.

It is clear from \textcolor{blue}{Fig. 2} that the radiation of the derived polarization is in good agreement with the full wave simulation. It is noteworthy that the small difference between the scattering patterns is due to the truncation of the scattering series and the dipole approximation.
 \begin{figure}[t!]
	\centering
	\epsfig{file=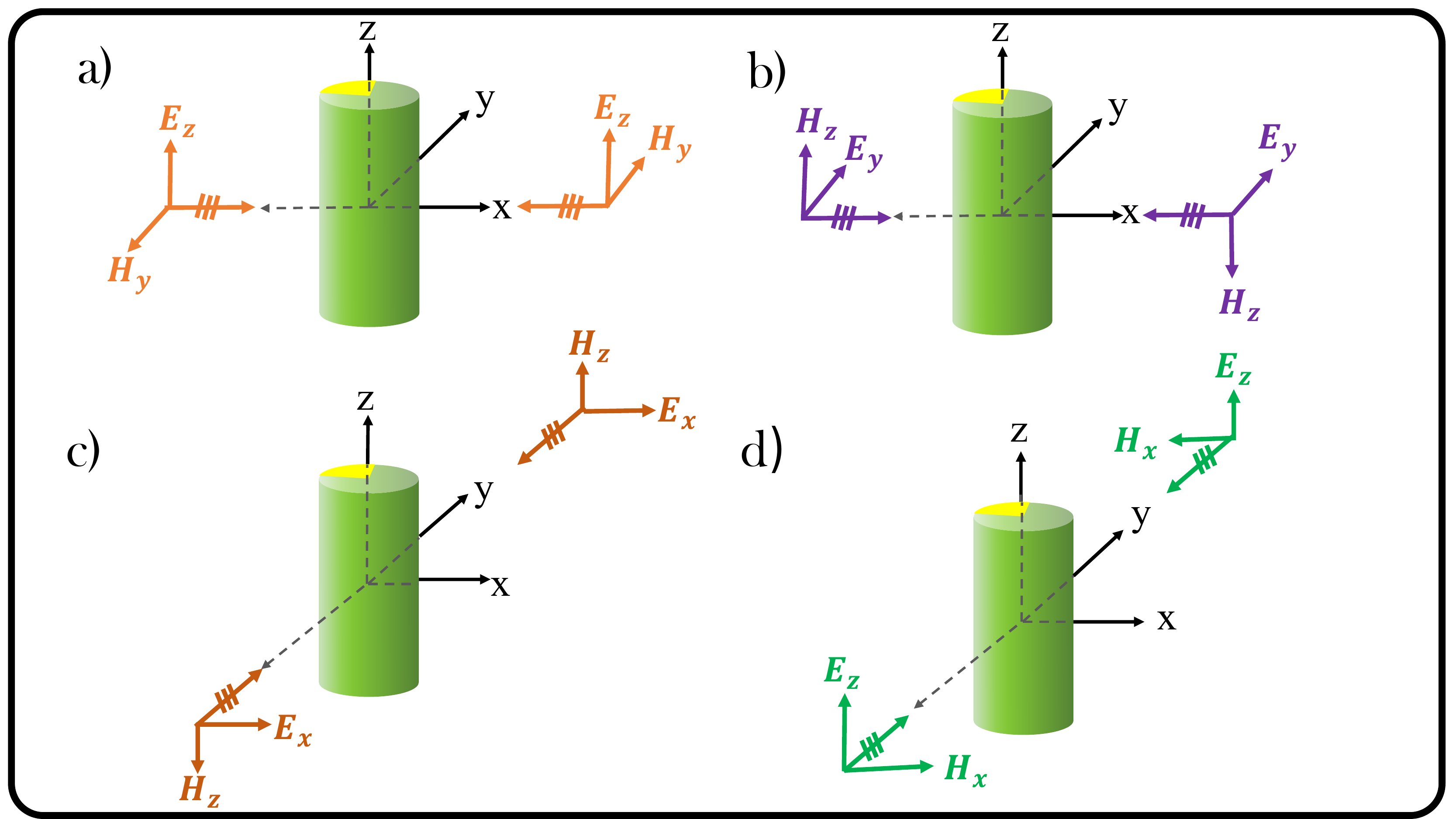, width=1\linewidth}  
	\caption{Polarizability tensor extraction setup based on standing wave method introduced in \cite{Asadchy,Yazdi1}. In each setup, we can calculate the a) $\alpha^{ee}_{zz}$, $\alpha^{me}_{yz}$, $\alpha^{em}_{zy}$, $\alpha^{mm}_{yy}$, $\alpha^{mm}_{xy}$, b) $\alpha^{ee}_{yy}$, $\alpha^{ee}_{xy}$, $\alpha^{me}_{zy}$, $\alpha^{em}_{yz}$, $\alpha^{em}_{xz}$, $\alpha^{mm}_{zz}$,  c) $\alpha^{ee}_{xx}$, $\alpha^{ee}_{yx}$, $\alpha^{me}_{zx}$, and d) $\alpha^{me}_{xz}$, $\alpha^{em}_{zx}$, $\alpha^{mm}_{xx}$, $\alpha^{mm}_{yx}$ individual polarizability components.  }\label{geom}
\end{figure}

\begin{figure}[t]
	\centering
	\centering\includegraphics[width=8cm]{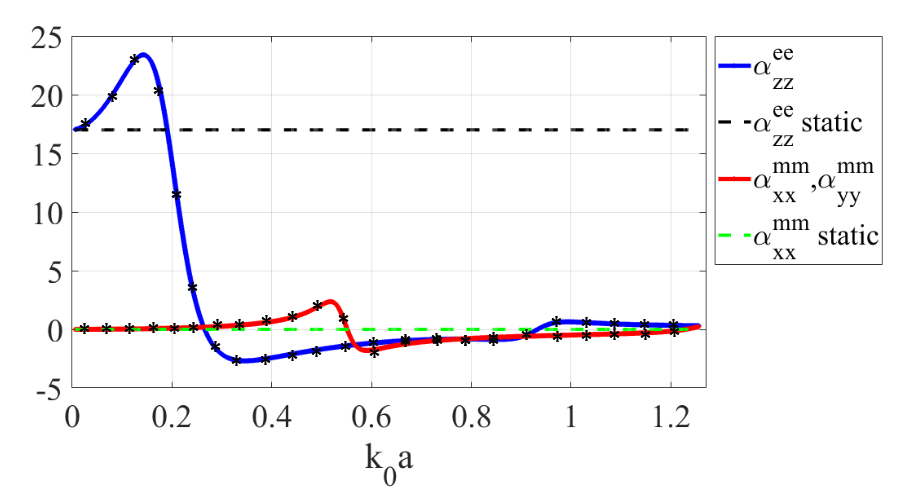} 
	\caption{Static and dynamic normalized transverse magnetic and longitudinal electric polarizability which exited by TM polarized incident wave for dielectric cylinder ($\epsilon_r=18$). The * markets are illustrated to verify the validity of method based on \cite{Strickland,Kallos}.}
\end{figure}
\subsection{Extraction of polarizability tensor}  

~~~Due to the asymmetry of the proposed meta-atom, in contrast to the symmetric cases of the sphere and the cylinder meta-atoms, extraction of the polarizability tensor is complicated. There are several methods to extract the polarizability tensor for a desired bianisotropic meta-atom \cite{Mirmoosa,Yazdi,Yazdi1,Asadchy}. In this paper, we use the standing wave approach to derive the polarizability tensor as presented in recent studies \cite{Yazdi1,Asadchy}. We can simply write a standing wave as a superposition of two plane waves traveling in opposite directions (see \textcolor{blue}{Fig.3}).  Using the extracted EM multiple moments from the previous section, the polarizability tensors excited by the standing wave can be derived using the superposition theorem. As it is clear from \textcolor{blue}{Fig. 3} the magnetic fields are out of phase in the center of coordinates, therefore by adding or subtracting the two plane waves we can derive the pure electric and magnetic components of the polarizability tensors, respectively. As an example 12 polarizability tensor components can be derived using ${E_x} = {E_0}{e^{ \pm jky}}$ and ${H_z} =  \pm {H_0}{e^{ \pm jky}}$  plane waves. 
\begin{align}
&   \alpha_{qx}^{ee}=\frac{P_{q}^{3\pi /2}+P_{q}^{\pi /2}}{2{{E}_{0}}},\alpha _{qx}^{me}=\frac{m_{q}^{3\pi /2}+m_{q}^{\pi /2}}{2{{E}_{0}}}\\
& \alpha_{qz}^{em}=\frac{P_{q}^{3\pi /2}-P_{q}^{\pi /2}}{2{{E}_{0}}}\eta,\alpha _{qz}^{mm}=\frac{m_{q}^{3\pi /2}-m_{q}^{\pi/ 2}}{2{{E}_{0}}}\eta 
\end{align}

where $q=\{x,y,z\}\nonumber$, and moments superscripts (i.e., $\pi/2$ and $3\pi/2$) indicate the wave vector angle with respect to the direction x. All the other components can be derived by simply using different configurations for the standing wave (see \textcolor{blue}{Fig. 3}) similar to \cite{Asadchy,Yazdi1}.

After complete consideration of all the derived polarizability tensor components, the complete polarizability tensor can be presented as follows:
\begin{widetext}
	\begin{align}{
	\begin{bmatrix}
	P       \\\\
	M      
	\end{bmatrix}
	=
	\begin{bmatrix}
	~~\alpha _{xx}^{ee} &~~\alpha _{xy}^{ee} & ~~0 &~~0  & ~~0 & ~~\alpha _{xz}^{em} ~~ 
	\\
	~~\alpha _{yx}^{ee} &~~\alpha _{yy}^{ee} &~~ 0 &~~0  & ~~0 &~~ \alpha _{yz}^{em} ~~\\
	~~0 & ~~0&~~ \alpha _{zz}^{ee} &~~\alpha _{zx}^{em}  &~~ \alpha _{zy}^{em} & 0~~\\
	~~0 & ~~0& ~~\alpha _{xz}^{me} &~~\alpha _{xx}^{mm}  &~~ \alpha _{xy}^{mm} & 0~~\\
	~~0 &~~ 0&~~ \alpha _{yz}^{me} &~~\alpha _{yx}^{mm}  &~~ \alpha _{yy}^{mm} & 0~~\\
	~~\alpha _{zx}^{me} &~~\alpha _{zy}^{me} &~~ 0 &~~0  & ~~0 & ~~\alpha _{zz}^{mm} ~~\\
	\end{bmatrix}
	\begin{bmatrix} E_{\text{loc}}       \\
	H_{\text{loc}}
	\end{bmatrix}	}
\end{align}
\end{widetext}
One should note that an analytical and compact form of the polarizability tensor components is extracted and presented in the \textbf{Appendix C}.

It is self-evident from the above relations that the induced polarizability tensor components depend on the Mie coefficients (wedge angle and dielectric constitutive parameters) and the meta-atom orientation, which make this meta-atom an excellent candidate for engineering and tuning of its EM response.
The present analysis has allowed us to define a fully dynamic expression for the entire polarizability tensor of the metal-dielectric meta-cylinder in closed-form. In the following section, we discuss the properties of this tensor and present some practical examples.

\section{PROPERTIES OF THE POLARIZABILITY TENSOR}
\begin{figure*}[t!]
	\centering
	\centering\includegraphics[width=13cm]{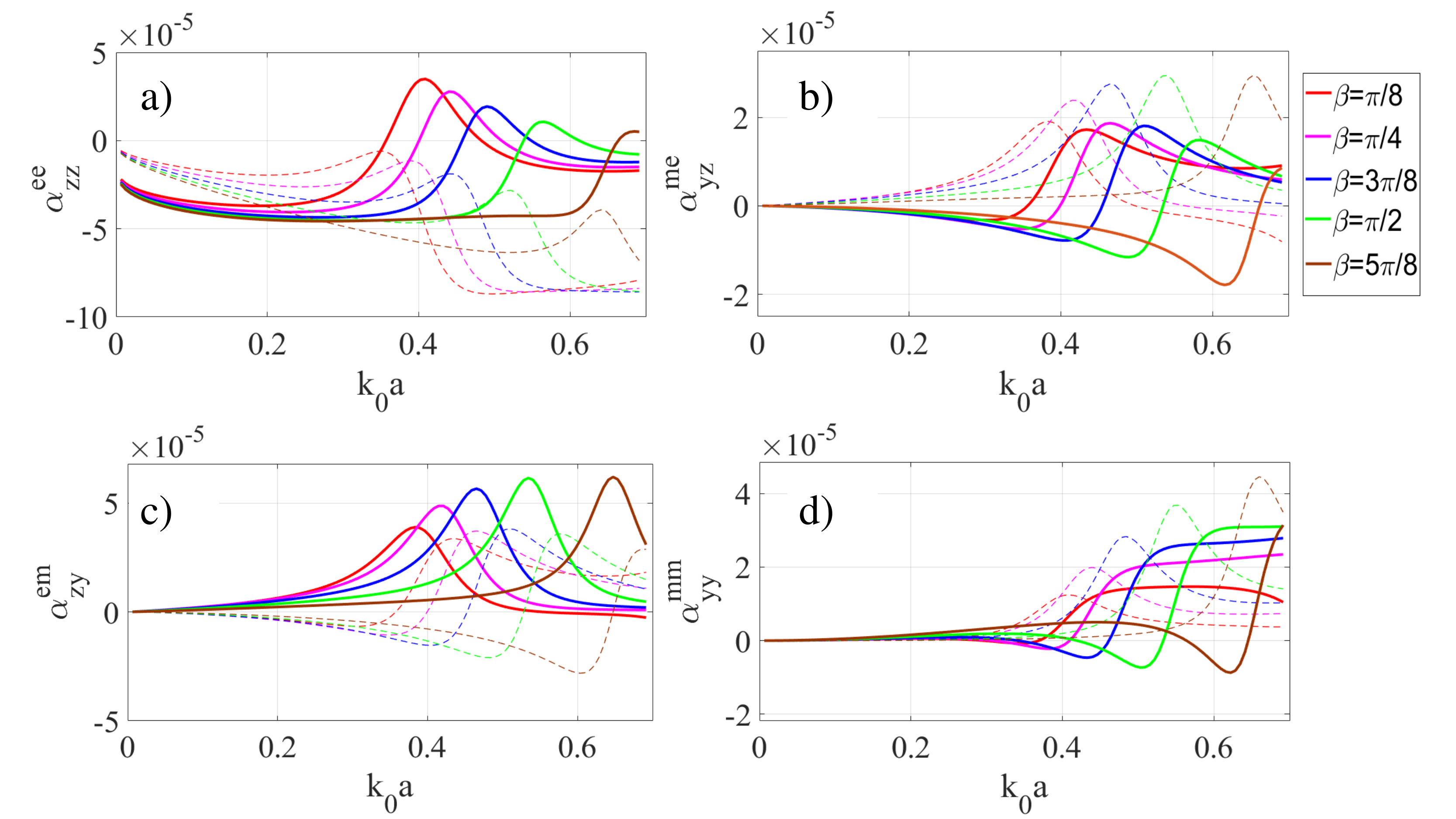}  
	\caption{ The real (solid lines) and imaginary parts (dashed lines) of a) electric, b) and c) magnetoelectric, and d) magnetic normalized dipole polarizability of the asymmetric metal-dielectric meta-cylinder (${\varepsilon }_{r}=18$) which exited by TM polarized incident wave for variant values of $\beta$ with respect to radius.}
\end{figure*}
~~~The polarizability tensor extracted in the previous section provides a fully dynamic, compact description of the EM response of a metal-dielectric meta-cylinder excited by an arbitrary electromagnetic wave. As noted, the validity of the tensor is limited to the case where the meta-atom is smaller than the wavelength.  Also, it’s scattering response can be approximated by using only the electric and magnetic dipole terms.
Although our derivation considered an infinitely long metal-dielectric meta-cylinder, the Mie scattering results are well-known to characterize a finite meta-atom with a length to diameter ratio of greater than 5 \cite{Kallos}.

We verify our results by simply comparing the polarizability components for the special case of an infinitely long  dielectric cylinder  ($\beta=0$) in both static \cite{Kallos} and dynamic regimes  \cite{Strickland}. 
\textcolor{blue}{Fig. 4} shows that the normalized transverse magnetic ($\alpha _{xx,yy}^{mm}/(\pi {{r_0}^{2}})$) and longitudinal electric polarizabilities ($\alpha _{zz}^{ee}/(\pi {{r_0}^{2}})$) are in good agreement with results of \cite{Strickland,Kallos}. 

The overarching objective of developing a tunable meta-atom is to enable the design of  any desired EM response, that is, to engineer a meta-atom for any given polarizability tensor. To-date, there are only a handful of meta-atoms represented by a closed-form EM response \cite{Basab,Strickland}. Further, it is noteworthy that the EM response of these meta-atoms is limited owing to the simplicity and symmetry of their structures. However, the metal-dielectric meta-cylinder presented here is a novel meta-atom capable of realizing complex EM responses through the tunability of its polarizability tensor.
The geometry of the presented meta-atom clearly shows that we have several degrees of freedom. We can simply tune the EM response and polarizability tensor by changing the constitutive and geometric parameters such as dielectric properties, wedge angle, and angle of rotation. These parameters affect the resonant frequency and the strength of the induced magnetic, electric and magnetoelectric responses. 

\begin{figure*}[t!]
	\centering
	\centering\includegraphics[width=13cm]{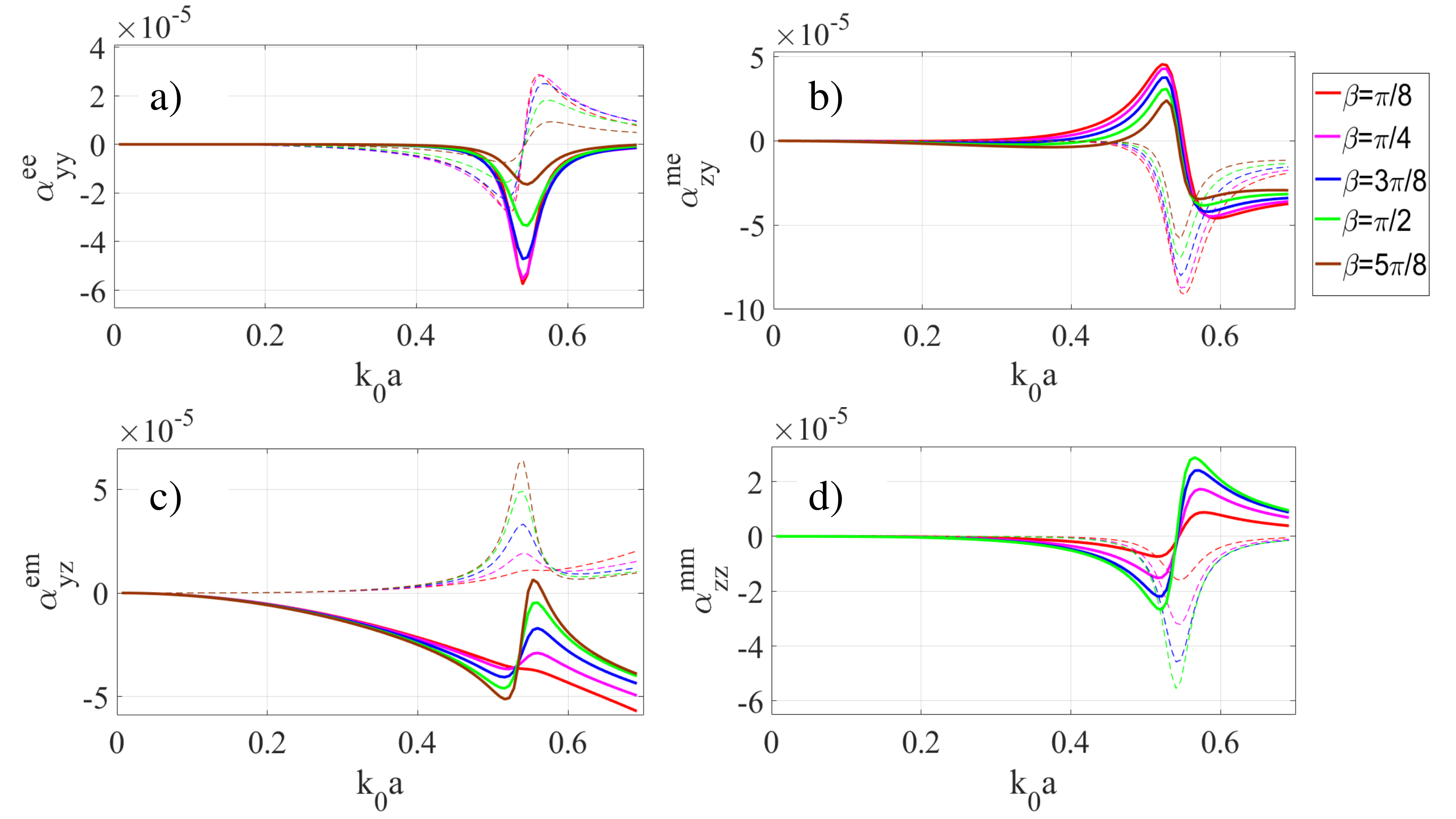} 
	\caption{ The real (solid lines) and imaginary parts (dashed lines) of a) electric, b) and c) magnetoelectric, and d: magnetic normalized dipole polarizability of the asymmetric metal-dielectric meta-cylinder (${\varepsilon }_{r}=18$) which is exited by TE polarized incident wave for variant values of $\beta$ with respect to radius.}
\end{figure*}
Since the tunability of the dielectric permittivity is limited to existing materials, the wedge angle  plays an important role in the design of the meta-atom. \textcolor{blue}{Figs. 5 and 6} illustrate the effect of wedge angle on the strength and position of the resonance with respect to the meta-cylinder radius. Here, we choose to illustrate a set of selected polarizability components for conciseness. Examining  \textcolor{blue}{Figs. 5 and 6}, we see that the resonant frequency of  the polarizability components excited by the TM wave can be tuned by adjusting the $\beta$ angle. However, adjusting the  wedge angle only affects the strength of the resonance in the TE excitation. This result is due to  the use of a non-magnetic dielectric in the meta-atom. Nevertheless, it is possible to simultaneously adjust both the strength and position of resonance with wedge angle when using a magnetic dielectric in the meta-atom. 
The presence of magnetoelectric effects in such a simple structure might be surprising at first sight. In fact, the geometrical asymmetry and inhomogeneity of this meta-atom breaks the symmetry in EM response and allows for the magnetoelectric effects to arise. One can also deduce this result from the standing wave approach. The EM response of an asymmetric structure is dissimilar for incident waves with opposite propagation directions, that is, the superposition of induced magnetic dipoles (see \textcolor{blue}{Eqs. 11} and \textcolor{blue}{12}) give rise to magnetoelectric polarizability components. 
\section{illustrative Applications}

~~~\textcolor{black}{By controlling the proposed nanostructure's  Mie resonance modes, we can attain a large number of nontrivial effects. Design of asymmetrical meta-atoms avail numerous applications in photonics. Breaking the geometrical symmetry leads to a high degree of freedom for engineering unconventional EM responses. In the following, we prove that the proposed meta-atom breaks azimuthal symmetry by reducing the coefficient associated with $sin (n\phi)$ modes to zero \cite{kivshar}. Based on our notation this condition leads to the following equation for electric dipole modes under TE illumination:
	\begin{align} &	\bigg| \bigg(a_{1}^{TE}-jb_{1}^{TE}\bigg)\cos (\alpha )+\bigg(b_{1}^{TE}+ja_{1}^{TE}\bigg)\sin (\alpha )\bigg|\neq \\
	&\bigg| \bigg(a_{1}^{TE}+jb_{1}^{TE}\bigg)\cos (\alpha )+\bigg(b_{1}^{TE}-ja_{1}^{TE}\bigg)\sin (\alpha )\bigg|	\end{align} 
	~The inequality in the above equation which be can  tune by changing $\alpha$,   is clearly the result of breaking the azimuthal symmetry in the system. This condition can lead to an unbalanced power radiation. Therefore, the proposed meta-atom can be designed (by adjusting $\alpha$) to achieve a desired unbalanced scattering pattern for beam deflection applications.   }
\\ 
\textcolor{black}{
	~In fact, breaking the azimuthal symmetry and attaining normal polarizability components are essential for design of novel meta-atomic structures used for various applications \cite{momeni2,momeni3,kivshar}. \\
}
Hereunder, we divide the applications of the proposed meta-cylinder to two different categories. 
\\1) \textbf{The microscopic level applications}:
Engineering the EM wave scattering of a single meta-atom, in which a dipole free response is desired. \\2) \textbf{The macroscopic level application}: Design of a metsurface processor consisting of densely packed meta-cylinders for optical signal processing and image processing (i.e., edge detection).

\subsection{Meta-atom: Near-zero scattering}
~~~First, we look for a condition where the electric polarization \textbf{\textit{P}} in the meta-atom can be made negligible (electric dipole-free meta-atom). We assume that the meta-atom is illuminated by a TM-polarized plane wave propagating in the x direction. According to \textcolor{blue}{Eq. 11} the electric polarization \textbf{\textit{$P_z$}} directly depends on the first harmonic of Mie scattering ($a_0$) of the metal-dielectric meta-cylinder. Targeting $a_0=0$ as a goal of the optimization,  we obtain the conditions under which the induced electric dipole is nearly zero ($\varepsilon_r =11.16,r=0.09\lambda ,\alpha =62.96^{\circ},\beta =125.27^{\circ}$). \textcolor{blue}{Fig. 7 a} shows the magnitude of electric polarization $P_z$  and transverse magnetic polarization $M_t$. Under this condition electric quadrupoles and magnetic dipoles entirely determine the wave-matter interaction which is illustrated through the 3D scattering pattern of the optimized meta-atom. It is worth noting that it is not possible to achieve a precisely zero value for the electric dipole in a passive structure. However, a recent study shows that this condition can be achieved when using a meta-atom structure which on the whole is lossless, that is, a structure comprising a loss-compensated dimer \cite{safari}.  
\begin{figure*}[t!]
	\centering
	\centering\includegraphics[width=18cm]{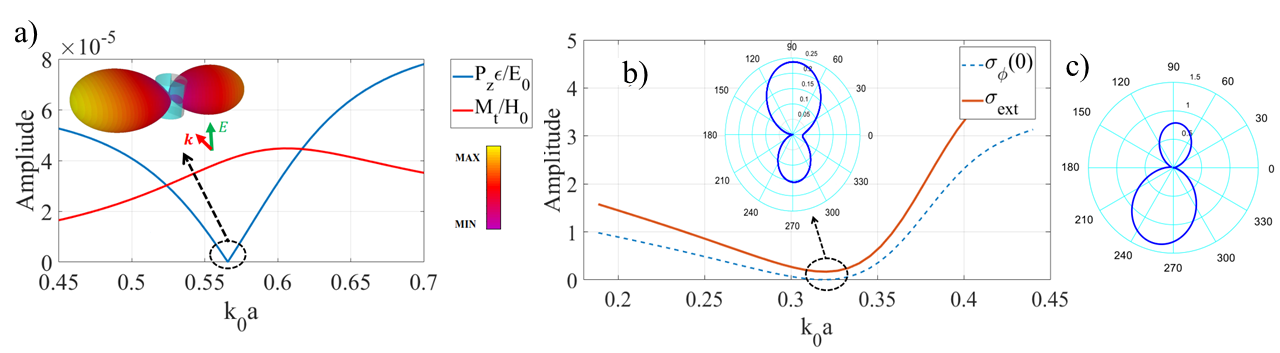}  
	\caption{a) Magnitude of electric polarization $P_z$ and transverse magnetic polarization $M_t$ with respect to radius and 3D scattering pattern for optimized electric dipole-free meta-atom. b) Total extinction cross section ($\sigma_{\text{ext}}$) and forward scattering ($\sigma_{\varphi}(0)$) with respect to radius and polar scattering pattern of the optimized meta-atom for near-zero forward scattering and c) backward scattering. }
\end{figure*}
~~~\textcolor{black}{Another appealing application is to determine the condition under which forward and backward scattering of the meta-atom vanishes. }
Based on the optical theorem \cite{Alu} the total extinction cross section of an object $\sigma_{ext}$  (sum of absorption and total scattering cross sections) is related to the normalized scattering amplitude in the forward direction  ${{\sigma }_{\varphi }}(0)$ in the following way
\begin{align}
{{\sigma }_{\text{ext}}}=\frac{\lambda _{0}^{2}}{\pi }\operatorname{Im}[{{\sigma }_{\varphi }}(0)]
\end{align} 
where $\lambda_0$ is the operating wavelength. Optical theorem applies to the EM response of any object illuminated by a linearly polarized plane wave. \textcolor{blue}{Eq. 18} implies that a near-zero forward scattering amplitude results in zero total scattering, which means that a meta-atom with near zero forward scattering will be transparent. Recent studies have demonstrated that in order for forward scattering to be zero, part of the meta-atom must be active. Therefore, in the passive case it is not possible to achieve exactly zero forward scattering \cite{Alu, safari}. 
Assuming that the meta-atom does not consist of absorptive material, the extinction cross section and scattering cross section would be equal. Hence, the total cross section can be calculated using the scattering fields in the following way. 
\begin{align}
& {{\sigma }_{T}}=\frac{1}{2\pi }\int\limits_{0}^{2\pi }{\sigma (\varphi )d\varphi } \\ 
& \frac{{{\sigma }_{T}}}{4r_0}=\frac{1}{2{{k}_{0}}r_0}\left\{ \left. \sum\limits_{n=0}^{\infty }{|a_{n}^{\text{sca}}{{|}^{2}}{{\varepsilon }_{n}}+\sum\limits_{n=1}^{\infty }{|b_{n}^{\text{sca}}{{|}^{2}}}} \right\} \right.
\end{align}

Now that we have the analytical expression for EM scattering from the proposed meta-atom, we can easily find the condition under which the forward and backward scattering are near-zero and zero, respectively. \textcolor{blue}{Figs. 7 b and c} present the near-zero/zero forward and backward scattering pattern under these specific conditions.

Here, the specific properties of the structure which possess near-zero/zero forward and backward scattering are presented. The electrical permittivity of the dielectric, radius of the cylinder, rotation angle, and wedge angle of a meta-atom with near-zero forward scattering are as follows:  ($\varepsilon_r =17.5,r=0.0507\lambda ,\alpha =90.01^\circ,\beta ={0.02^\circ}$) . \textcolor{blue}{Fig. 7 b} illustrates the 2D scattering of the meta-atom with the aforementioned properties and the total scattering with respect to wavelength. A similar condition for zero backscattering is found ($\varepsilon_r =14.807,r=0.0721\lambda ,\alpha =279.87^\circ,\beta ={51.36^\circ}$) and the corresponding scattering pattern is shown in \textcolor{blue}{Fig. 7 b}.  \textcolor{blue}{Fig 7 b} shows that the forward scattering and the total cross section are highly correlated and that the minimum total scattering and the forward scattering occur at the same wavelength.

\subsection{Metasurface Processor:\\ Advanced Spatial Optical Signal Processing}
\textcolor{black}{Preforming} wave-based spatial optical signal and image processing with metasurface processors has various advantages such as compact single block architecture, high speed response, and low energy consumption compared to the conventional electronic-based devices \cite{kwon2020dual,kwon2018nonlocal,lou2020optical}. 
One of the main challenges of optical signal processing today, is to design an optical spatial filter with an asymmetrical (i.e., odd) OTF response capable of distinguishing between $-k_t$ and $k_t$ propagation at normal incidence. The asymmetrical OTF response is
required for first order differentiation using metasurface processors \cite{momeni2,momeni3,cordaro2019high,kwon2018nonlocal,zhu2019generalized,lou2020optical}. Herein, we discuss the ability of the proposed meta-atom to engineer asymmetrical OTF responses by virtue of tuning the wedge angle and the dielectric constant making this structure amenable to processing applications such as first order differentiation.
Consider a  homogeneous passive reciprocal metasurface layer comprising the proposed meta-cylinders located at y=0 plane, 
where, $\psi _{\text{inc}}({{x}})$ and $\psi _{\text{ref/tran}}({{x}})$ are the input and output signals of this linear optical system (see \textcolor{blue}{Fig. 8 a}). The angular EM response of the metasurface processor determines its' corresponding OTF response in the spatial Fourier domain (i.e., $\tilde{H}( {{k}_{x}})$). 
The output signal $\psi _{\text{inc}}({{x}})$ can be calculated given the input signal $\psi _{\text{ref/tran}}({{x}})$ using ${{\psi}_{\text{ref/tran}}}\left( x \right)={{F}^{-1}}\left[ \tilde{H}\left( {{k}_{x}} \right)\times \\
F\left( {{\psi}_{inc}}\left( x \right) \right) \right]$, where $k_{x}$ denotes the spatial frequency variable in the Fourier space, $F$ and $F^{-1}$ represent the Fourier and inverse Fourier transforms, respectively. 
Here, we first analyze a metasurface processor, that is an array of electric/magnetic dipoles comprising both tangential (t-subscript) and normal (n-subscript) components. 
\begin{figure*}[t!]
	\centering
	\centering\includegraphics[width=16cm]{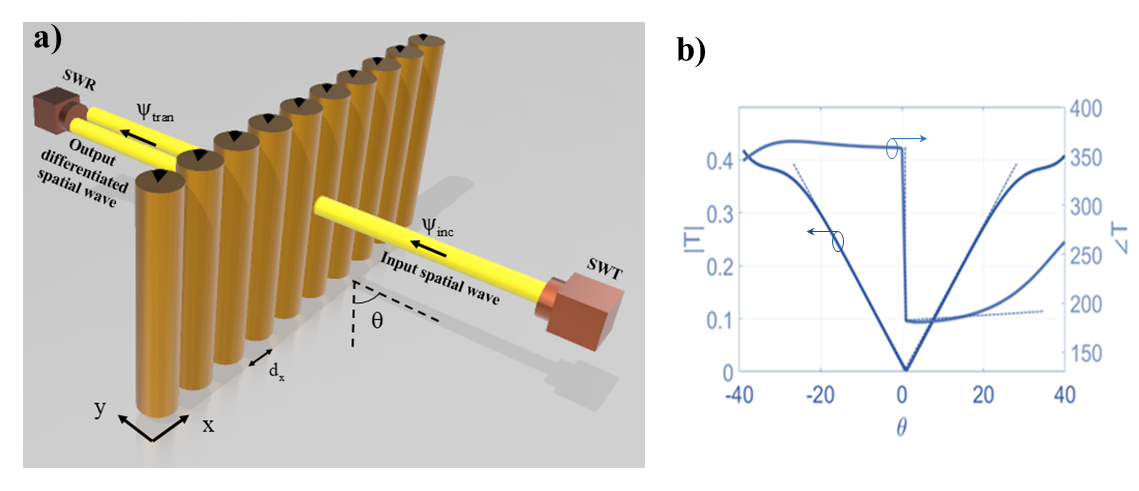} 
	\caption{a) Array of metal-dielectric meta-cylinders (include silicon and gold \cite{palik1998handbook} materials ) as metasurface processor. SWT and SWR are spatial wave transmitter and receiver, respectively. b) solid line: the amplitude and phase of synthesized OTF Dashed line: The ideal OTF for the first-order differentiation
		operation. Simulation parameters: $\lambda_0=4.16 \mu m$,  $r_0=0.9 \mu m$, $\beta=60^\circ$ $\alpha=150^\circ$, and $d_x=2.5 \mu m$}
\end{figure*}
The spatial transfer function associated with the metasurface processor can be modeled using the vectorial form of the generalized
sheet boundary conditions \cite{albooyeh2017equivalent}.
\begin{align}
&\textbf{E}_{\rm{t}}^ +  \times {\rm{\hat{ \textbf{ n}}}} - \textbf{E}_{\rm{t}}^ -  \times {\rm{\hat{ \textbf{ n}}}} =  - i\omega \bigg({\textbf{M}_{\rm{t}}} - {\rm{\hat{ \textbf{ n}}}} \times \frac{{{\textbf{k}_\text{t}}}}{{\omega \varepsilon }}{P_{\rm{n}}}\bigg),\ \\ 
&{\rm{\hat n}} \times {\rm{\textbf{H}}}_{\rm{t}}^ +  - {\rm{\hat {\textbf{n}}}} \times {\rm{\textbf{H}}}_{\rm{t}}^ -  =  - i\omega \bigg({{\rm{\textbf{P}}}_{\rm{t}}} + {\rm{\hat {\textbf{n}}}} \times \frac{{{{\rm{\textbf{k}}}_t}}}{{\omega \mu }}{M_{\rm{n}}}\bigg),\
\end{align}
where $\epsilon$ and $\mu$ are the permittivity and permeability of
the surrounding medium, respectively. ${\hat n}$ is the unity vector normal to the metasurface plane (i.e.,  ${\hat y}$). The + and - superscripts refer to the field values
at $ y = 0^+$ and $0^-$, respectively. Moreover, t and n subscripts represent the tangential and normal components. In order to simplify the eqs. 21 and 22 we can introduce the equivalent magnetic and electric surface polarization densities in the following way
\begin{align}
& {{\rm{\textbf{M}}}_{{\rm{t,eq}}}}{\rm{ = }}{{\rm{\textbf{M}}}_{\rm{t}}} - {\rm{\hat n \times }}\frac{{{{\rm{\textbf{k}}}_{\rm{t}}}}}{{\omega \varepsilon }}{P_n}\, \\ 
&{{\rm{\textbf{P}}}_{{\rm{t,eq}}}}{\rm{ = }}{{\rm{\textbf{P}}}_{\rm{t}}} + {\rm{\hat n \times }}\frac{{{{\rm{\textbf{k}}}_{\rm{t}}}}}{{\omega \mu }}{M_n}\  \nonumber
\end{align}
Furthermore, the tangential components of the reflected $E^r_t$ and
transmitted $E^t_t$ electric fields at the boundary of the metasurface are obtained
\begin{align}
&{\rm{\textbf{E}}}_{\rm{t}}^{\rm{r}}{\rm{ = }}\frac{{i\omega }}{2}{\rm Q}.\bigg[{{\rm{\textbf{P}}}_{{\rm{t,eq}}}} \mp {\rm Q^{ - 1}}.\hat {\textbf{n}} \times {{\rm{\textbf{M}}}_{{\rm{t,eq}}}}\bigg]\\\
&{\rm{\textbf{E}}}_{\rm{t}}^{\rm{t}}{\rm{ = \textbf{E}}}_{\rm{t}}^{\rm{i}} + \frac{{i\omega }}{2}{\rm Q}.\bigg[{{\rm{\textbf{P}}}_{{\rm{t,eq}}}} \pm {\rm Q^{ - 1}}. \hat {\textbf{n}} \times {{\rm{\textbf{M}}}_{{\rm{t,eq}}}}\bigg]\ 
\end{align}
The derivation of $Q$, $P_t$, and $P_n$ is outlined in the \textbf{Appendix D} . 
To further simplify the equations  $\bold{M_t}=\bold{M_n}=0$ can be considered for non-magnetic materials. Thereafter, reflection and transmission coefficients are derived \begin{align}
{\rm{R = }}\frac{{j\omega \eta }}{{{\rm{2S}}}}\bigg[\hat \alpha _{{\rm{xx}}}^{{\rm{ee}}}|cos\theta | - \frac{{\hat \alpha _{{\rm{yy}}}^{{\rm{ee}}}{{\sin }^2}\theta }}{{|cos\theta |}}\bigg],\
\end{align}
\begin{align}
{\rm{T = }}\frac{{{\rm{j}}\omega \eta }}{{{\rm{2S}}}}\bigg[\hat \alpha _{{\rm{xx}}}^{{\rm{ee}}}|cos\theta | \pm 2\hat \alpha _{{\rm{xy}}}^{{\rm{ee}}}\sin \theta  + \frac{{\hat \alpha _{yy}^{{\rm{ee}}}{{\sin }^2}\theta }}{{|cos\theta |}}\bigg],\
\end{align}
where $\hat \alpha ^{ee}$ is the effective polarizability. Upon closer observation of the equation (27), it is clear that an odd spatial OTF response (i.e., $sin(\theta)$ angular variation) can only be achieved by exciting the $\hat \alpha _{\text{ee}}^{xy}$ component on the metasurface processor. As we mentioned earlier in this section, achieving an odd OTF response is required for first order differentiation which highlights ability of the proposed met-atom to introduce \textcolor{black}{normal} polarizability components to the metasurface processor. 
Although the analytical formula for the individual polarizability of the meta-cylinder is described here, the analytical derivation of effective polarizability tensor from individual polarizability tensor  is well studied \cite{laroche2006tuning,niemi2013synthesis}.

Given the fact that the excitation of xy component of the effective polarizability necessitates the excitation of the respective component of the individual polarizability tensor, excitation of xy component in the  individual polarizability is required for breaking the angular symmetry of transmission response (i.e., distinguishing between $k_x$ and $-k_x$). In the section III, we explicitly show that the proposed meta-atom is well capable of engineering xy polariszability component: 
\begin{align}
\alpha _{xy}^{ee} =  - \frac{{2\left( {\Psi _{b,1}^{TE, + }\cos (\alpha ) - \Psi _{a,1}^{TE, + }\sin (\alpha )} \right)}}{{jk_0^2}}
\end{align} where the value of $\alpha_{xy}^{ee}$ can be engineered by virtue of tuning the geometrical properties of the meta-atom (i.e., $\beta$ and $\alpha$). \\
~~ Now that the xy polazibality can be achieved using the proposed meta-cylinder, we  optimize the optical response of the metasurface processor comprising an array of meta-cylinders to achieve the metasurface differentiator's OTF response, that is $H(k_x) = c jk_x = cjk \sin\theta$, where c is a constant. This first-order differentiation operator is of high importance for various applications in signal and image processing such as edge detection. Comparing the mentioned OTF response with eq. 27 the optimization goal for the desired metasurface processor is derived: $\alpha_{xx}^{ee}=\alpha_{yy}^{ee}=0$ and for $-30^\circ<\theta<30^\circ$. 

\begin{figure*}[t!]
	\centering
	\centering\includegraphics[width=12cm]{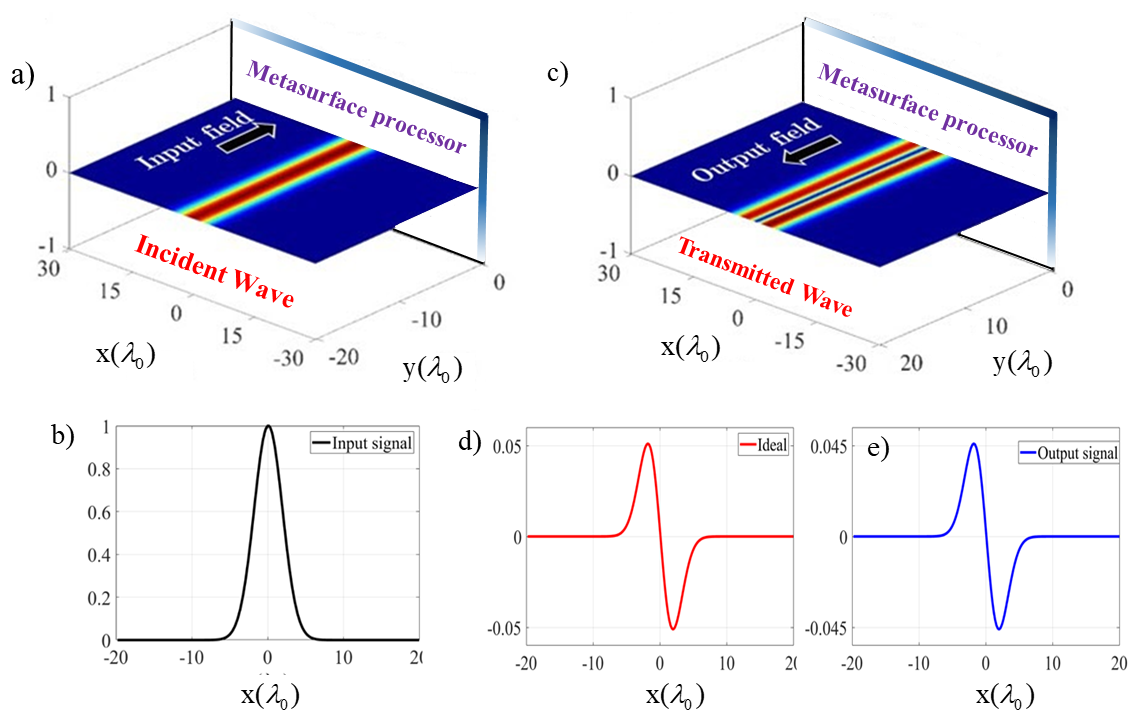}  
	\caption{a), b) The Gaussian-shape incident field profile with c), e) the transmitted derivative field profile. d) The exact derivative
		signals have been also presented for the sake of comparison.}
\end{figure*}
\begin{figure}[t!]
	\centering
	\centering\includegraphics[width=8cm]{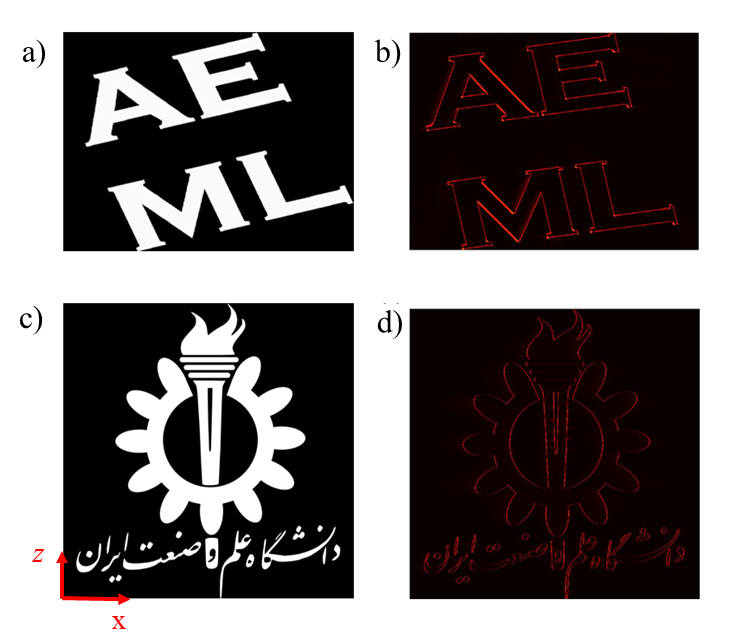}  
	\caption{ Edge Detection of images by exploiting the proposed metasurface processor. a) and c) Input images and b) and d) Edge-detected images when metasurface differentiator differentiates the input image along x direction. }
\end{figure}
Here, the geometry of the proposed meta-atom gives us the ability us to achieve this goal and synthesize the desired OTF response which is enables real-time first-order derivation (see \textcolor{blue}{Fig. 8}).
The proposed meta-atom can elaborately mimic the required $k_{x}$-dependency first-order differentiator (see \textcolor{blue}{Fig. 8b}). 
An excellent agreement between the transmission coefficient of the designed metasurface and the the desired OTF response for both phase and amplitude has been achieved for input signals with the normalized spectral beamwidth $|W/k_0|< 0.3$.

To verify the validity of the proposed metasurface differentiator performance of the employed metasurface processor, a Gaussian signal is considered as the input signals (see \textcolor{blue}{Figs. 9 a, b}). The transmitted signal
and the corresponding results are illustrated in \textcolor{blue}{Figs. 9 c, d, and e}. To evaluate the performance of our metasurface differentiator, the numerically-obtained result is
compared with the first-derivative of the input signal (\textcolor{blue}{see Fig. 9 d}).
An excellent agreement between output and first derivative has been achieved.\\

As it was mentioned earlier, optical metasurface operators are highly advantages in image processing. Here in particular, the designed differentiator can be used for edge detection which plays a momentous role in the image
segmentation and the other image pre-processing steps \cite{momeni2,momeni3}. To
elaborate even furthur, object recognition with the help of spatial differentiation
enables us to extract the boundaries between two regions of different intensities.

To show the ability of the designed metasurfacce processor, two images have been used for edge detection
(see \textcolor{blue}{Figs. 10 a and c}). The transmitted images
are numerically simulated and the corresponding transverse
field profiles are displayed on the right hand side of \textcolor{blue}{Figs. 10 b and d}.
As expected, the 1D edge-detector metasurface successfully
exposes all outlines of the normally incident image along the
vertical orientations, exhibiting its higher sensibility to fine
details as a first-order derivative operator. 
\section{Conclusion}
~~~An ideal meta-atom is one which contains all possible bianisotropic electromagnetic responses wherein the structure of the meta-atom permits independent tunability of all the elements comprising the constitutive parameters. Nevertheless, there will be some physical constraints to the design of any meta-atom. 
In this paper, we propose a novel metal-dielectric meta-cylinder as a tunable meta-atom which consists of metallic and dielectric finite wedges. Numerous design parameters provide a good degree of freedom which enable the  proposed meta-atom to be an excellent candidate for engineering the EM response. We first derive the analytical expression for the EM scattering due to the structure. Next, we extract the polarizability tensor and explaining the response of such a meta-atom in the dipole region. Finally, we show the flexibility of the proposed inclusion through practical applications in both microscopic and macroscopic levels. The examples verify the validity of the analytical formulation and illustrate the capability of the proposed meta-atom to independently tune the electromagnetic polarizability components.
\\
\appendix

\section{Boundary Conditions}
In this Appendix, we detail the analytical steps for deriving
EM scattering from an asymmetric metal-dielectric meta-cylinder of infinite extent along its axis. \\

In the dielectric region, we construct an expansion for the total electric field $E^{d}= E_z^d\hat z$ the TM case, which follows from the solution of Maxwell's equations in cylindrical coordinates
\begin{align}
E_{z}^{d}(r,\phi) =&\sum\limits_{i=0}^{\infty }\bigg(\zeta^\text{I}_{\nu i} J_{\nu i}(k_c r)\cos (\nu i (\varphi+\alpha)\bigg)+\\
&\sum\limits_{i=1}^{\infty }\bigg(\zeta^{\text{II}}_{\tau i} J_{\tau i}(k_c r)\sin (\tau i (\varphi+\alpha))\bigg)\nonumber
\end{align}
where $k_c=\omega \sqrt{\epsilon_c \mu_c}$ denotes the wavenumber. The $\zeta^\text{I}$ and the
$\zeta^\text{II}$ are the (unknown) coefficients. Also, the incident plane wave has a following form 
\begin{align}
E_{z}^{inc}(r,\phi) =&\sum\limits_{n=0}^{\infty }\bigg(U^{\text{I}}_{n} J_{n}(k_0 r)\cos (n( \varphi+\alpha))\bigg)+\\
&\sum\limits_{i=1}^{\infty }\bigg(U_{n}^{\text{II}} J_{\tau i}(k_0 r)\sin (n (\varphi+\alpha))\bigg)\nonumber
\end{align}
where 
\begin{align}
U^{\text{I}}_{n} = \frac{2E_0j^n}{\epsilon_n} \cos (n(\gamma+\alpha))\\ 
U^{\text{II}}_{n} = 2E_0j^n \sin (n(\gamma+\alpha)) 
\end{align}
and also $E_z^{sc}$ is presented in the equation (1). On the metal part of dielectric-metal meta-cylinder 
($r = r_0$) the tangential part of the total electric field must vanish, and at the dielectric-air interface ($r=r_0$), the tangential parts of the total electric and total magnetic field have to be continuous. 
\begin{align}
E_{z}^{sc}|_{r=r0}+E_{z}^{inc}|_{r=r0}=\   \left\{
\begin{array}{ll}
E_{z}^{d}|_{r=r0} & S_1<\phi<S_2\\
0  & \text{Otherwise}\nonumber
\end{array} 
\right. \\\
\end{align}
\begin{align}
H_{\phi}^{sc}|_{r=r0}+H_{\phi}^{inc}|_{r=r0}=H_{\phi}^{d}|_{r=r0} ~~~~~  S_1<\phi<S_2
\end{align}
where $S_1$ and $S_2$ are $-\pi+(\beta/2+\alpha)$ and $\pi-(\beta/2+\alpha)$. 

Using the orthogonality relations of the
trigonometric functions, we can compute the unknown scattering coefficient ($a_n$ and $b_n$) in equations (1) and (2) which are presented in equations (5) and (6). 

Also,  $\tau i$ ,$\nu i$,  $\varepsilon _n$ , and $\delta _{nm}$ are defined in the following way
\begin{align}
&\tau i=\frac{i\pi}{\pi-\beta/2} \,\,\,\,\,\,\,\,\,\,\,\, \,\,\,\,\,\,\,\,\,\,\,\,\, i=1,2,3,... \\
&\nu i=\frac{\pi(2i+1)}{2(\pi-\beta/2)} \,\,\,\,\,\,\,\,\,\, \,\,\,\,\,i=0,1,2,...\\
&{\varepsilon _n} = \left\{ {\begin{array}{*{20}{c}}
	{2{\rm{\,\,\,\,\,\,n  =  0}}}\\
	{1{\rm{\,\,\,\,\,\,n}} \ne {\rm{0}}}
	\end{array}} \right.\\
&{\delta _{nm}} = \left\{ {\begin{array}{*{20}{c}}
	{1{\rm{\,\,\,\,\,\,n  =  m}}}\\
	{0{\rm{\,\,\,\,\,\,n}} \ne {\rm{m}}}
	\end{array}} \right.\
\end{align}
\\
\section{Scattering Coefficients}
The scattering coefficients of metal-dielectric meta-atom $a_{n}$ and $b_{n}$ are derived using the EM boundary conditions.  Further, these can be written in the following way where 'e' and 'o' superscripts allude to the even and odd modes in the Mie scattering coefficient.
\begin{align}
&\bigg[a_{n}^{TM/TE}\bigg]_{n\times1}=\\
&\bigg[\tau_{n}\bigg]_{1\times n}\bigg[{\chi_{nm}}^{(e),TM/TE}\bigg]_{n\times m}^{-1}\bigg[\Upsilon_{m}^{(e),TM/TE}\bigg]_{m\times1}\nonumber
\end{align}
\begin{align}
&\bigg[b_{n}^{TM/TE}\bigg]_{n\times 1}=\\
&\bigg[\tau_{n}\bigg]_{1\times n}\bigg[\chi_{nm}^{(o),TM/TE}\bigg]^{-1}_{n\times m}{\bigg[\Upsilon _{m}^{(o),TM/TE}\bigg]_{m\times 1}}\nonumber
\end{align}
where the $\chi_{nm}^{(e),TM}$,${\tau}_{n}$, and $\Upsilon_{m}^{(e),TM}$ are given by:
\begin{align}
\tau_{n}=\frac{1}{\frac{dH_{n}^{{1}}({{k}_{0}}{{R}_{0}})}{dkr}} 
\end{align} 
\begin{align} 
&\chi _{nm}^{(e)}=\frac{H_{n}^{{1}}({{k}_{0}}{{r}_{0}})}{\frac{dH_{n}^{{}}({{k}_{0}}r)}{dk_0r}}\pi {{\varepsilon }_{n}}{{\delta }_{nm}}-\sum\limits_{i=0}^{\infty }{\frac{{{Z}_{c}}(\pi -\beta /2)}{{{Z}_{0}}}\frac{{{J}_{\nu i}}({{k}_{c}}{{r}_{0}})}{\frac{d{{J}_{\nu i}}({{k}_{c}}{{r}_{0}})}{dk_cr}}}\\
&\times \bigg(\text{sinc}(\frac{(\pi -\beta /2)(\nu i-n)}{\pi })+\text{sinc}(\frac{(\pi -\beta /2)(\nu i+n)}{\pi })\bigg)\\
&\times \bigg(\text{sinc}(\frac{(\pi -\beta /2)(\nu i-m)}{\pi })+\text{sinc}(\frac{(\pi -\beta /2)(\nu i+m)}{\pi })\bigg), \end{align}
\begin{align}&\Upsilon_{m}^{(e)}= -a_{m}^{\text{inc}}{{J}_{m}}({{k}_{0}}{{r}_{0}})\pi {{\varepsilon }_{m}}+\sum\limits_{i=0}^{\infty}
\frac{{{Z}_{c}}(\pi -\beta /2)}{{{Z}_{0}}}\frac{{{J}_{\nu i}}({{k}_{c}}{{r}_{0}})}{\frac{d{{J}_{\nu i}}({{k}_{c}}{{r}_{0}})}{dk_cr}}\\
&\times 
\bigg(\text{sinc}(\frac{(\pi -\beta /2)(\nu i-m)}{\pi })+\text{sinc}(\frac{(\pi -\beta /2)(\nu i+m)}{\pi })\bigg)\\
&\times\sum\limits_{n=0}^{\infty }a_{n}^{\text{inc}}\frac{d{{J}_{n}}({{k}_{0}}r)}{dk_0r}\bigg(\text{sinc}(\frac{(\pi -\beta /2)(\nu i-n)}{\pi })+\\
&\text{sinc}(\frac{(\pi -\beta /2)(\nu i+n)}{\pi })\bigg), \end{align}
The relations for $\chi _{nm}^{(o),TM}$ and $\Upsilon _{m}^{(o),TM}$ can be written similar to $\chi _{nm}^{(e),TM}$ and $\Upsilon _{m}^{(e),TM}$ using the following substitutions:  $\nu i\to \tau i$, ${{\varepsilon }_{n}}\to 1$, $a_{m}^{inc}\to b_{m}^{inc}$, $\text{sinc}({{\theta }_{1}})+\text{sinc}({{\theta }_{2}})\to \text{sinc}({{\theta }_{1}})-\text{sinc}({{\theta }_{2}})$, and $(i=n=0) \to (i=n=1)$ (indexes of the summation). Also the dual relations for $\chi _{nm}^{(o),TE}$ and $\Upsilon _{m}^{(o),TE}$ can be written with the following substitutions in  $\chi _{nm}^{(o),TM}$ and $\Upsilon _{m}^{(o),TM}$:  $\nu i\to \tau i$, $\tau i\to \nu i$, ${J_n}(\chi ),{H_n}(\chi ) \rightleftharpoons  \frac{{d{J_n}(\chi )}}{{d\chi }},\frac{{d{H_n}(\chi )}}{{d\chi }}$ and $\frac{{{Z}_{c}}(\pi -\beta /2)}{{{Z}_{0}}}\to \frac{{{Z}_{c}}(\pi -\beta /2)}{{{\varepsilon }_{i}}{{Z}_{0}}}$, where ${{Z}_{0}}$ and ${{Z}_{c}}$ are wave impedance of background medium and dielectric cylinder, respectively.

\section{Polarizability Tensor Components }

In order to present a compact expression for the extracted polarizability tensor components, we define the following arbitrary parameters which depend on the Mie coefficients: 
\begin{equation}
\begin{array}{l}
\psi _{{\rm{N}},{\rm{M}}}^{{\rm{R}}, \pm } = N_M^{R,2\pi } \pm \;N_M^{R,\pi }
\end{array}
\end{equation}
\begin{equation}
\begin{array}{l}
^\dag \psi _{{\rm{N}},{\rm{M}}}^{{\rm{R}}, \pm } = N_M^{R,3\pi /2} \pm \;N_M^{R,\pi /2}
\end{array}
\end{equation}
where R=TM or TE, N= a or b and M=0 or 1.
{We normalized all the presented polarizabilities to provide fair comparison: $\alpha_{\rm ee}$, $\alpha_{\rm em}$,$\alpha_{\rm me}$ and $\alpha_{\rm mm}$ are normalized respectively to $\epsilon_0$, $(\epsilon_0 Z_0)^{-1}$, $Z_0$ and 1. }  
Using the above expressions, the non-zero polarizability components are obtained as follows:  \\
\begin{widetext}
\begin{center}
	\begin{tabularx}{\textwidth}{ X X X}{\centering\arraybackslash}
		\\
		& \textbf{Polarizability tensor components} & \\
		\hline
		\\
		$ \alpha_{zz}^{ee}=\frac{2\Psi _{{\rm{a,0}}}^{{\rm{TM,+}}}}{jk_{0}^{2}} $  & $ \alpha_{yz}^{me} =-\frac{{\Psi _{{\rm{a,1}}}^{{\rm{TM, + }}}\cos (\alpha ) + \Psi _{{\rm{b,1}}}^{{\rm{TM, + }}}\sin (\alpha )}}{{jk_0^2}}$ & $\alpha _{zy}^{em} = \frac{{2\Psi _{a,0}^{TM, - }}}{{jk_0^2}}$  \\
		$\alpha _{yy}^{mm} =  - \frac{{\left( {\Psi _{a,1}^{TM, - }\cos (\alpha ) + \Psi _{b,1}^{TM, - }\sin (\alpha )} \right)}}{{jk_0^2}}$  & $\alpha _{xy}^{mm} = \frac{{\left( {\Psi _{b,1}^{TM, - }\cos (\alpha ) - \Psi _{a,1}^{TM, - }\sin (\alpha )} \right)}}{{jk_0^2}} $ & $\alpha _{yy}^{ee} = \frac{{2\left( {\Psi _{a,1}^{TE, + }\cos (\alpha ) + \Psi _{b,1}^{TE, - }\sin (\alpha )} \right)}}{{jk_0^2}}$    \\
		$\alpha _{xy}^{ee} =  - \frac{{2\left( {\Psi _{b,1}^{TE, + }\cos (\alpha ) - \Psi _{a,1}^{TE, + }\sin (\alpha )} \right)}}{{jk_0^2}}$  & $\alpha _{zy}^{me} = \frac{{2\Psi _{a,0}^{TE, + }}}{{jk_0^2}}$  & $\alpha _{yz}^{em} = \frac{{2\left( {\Psi _{a,1}^{TE, - }\cos (\alpha ) + \Psi _{b,1}^{TE, - }\sin (\alpha )} \right)}}{{jk_0^2}}$    \\
		$\alpha _{xz}^{em} =  - \frac{{2\left( {\Psi _{b,1}^{TE, - }\cos (\alpha ) - \Psi _{a,1}^{TE, - }\sin (\alpha )} \right)}}{{jk_0^2}}$  & $\alpha _{zz}^{mm}=\frac{{2\Psi _{a,0}^{TE, - }}}{{jk_0^2}}$  & $\alpha _{xx}^{ee} =  - \frac{{2\left( {{}^\dag \Psi _{b,1}^{TE, + }\cos (\alpha ) - {}^\dag \Psi _{a,1}^{TE, + }\sin (\alpha )} \right)}}{{jk_0^2}}$    \\
		$\alpha _{yx}^{ee} = \frac{{2\left( {{}^\dag \Psi _{a,1}^{TE, + }\cos (\alpha ) + {}^\dag \Psi _{b,1}^{TE, + }\sin (\alpha )} \right)}}{{jk_0^2}}$  & $\alpha _{zx}^{me} = \frac{{{2^\dag }\Psi _{a,0}^{TE, + }}}{{jk_0^2}}$  & $\alpha _{xz}^{me} = \frac{{\left( {^\dag \Psi _{b,1}^{TM, + }\cos (\alpha ){ - ^\dag }\Psi _{a,1}^{TM, + }\sin (\alpha )} \right)}}{{jk_0^2}}$    \\
		$\alpha _{zx}^{em} = \frac{{{2^\dag }\Psi _{a,0}^{TM, - }}}{{jk_0^2}}$  & $\alpha _{xx}^{mm} = \frac{{^\dag \Psi _{b,1}^{TM, - }\cos (\alpha ){ - ^\dag }\Psi _{a,1}^{TM, - }\sin (\alpha )}}{{jk_0^2}}$  & $\alpha _{yx}^{mm} =- \frac{{\left( {^\dag \Psi _{a,1}^{TM, - }\cos (\alpha ){ + ^\dag }\Psi _{b,1}^{TM, - }\sin (\alpha )} \right)}}{{jk_0^2}}$   \\
		\hline\\
	\end{tabularx}
\end{center}
\end{widetext}

\section{Effective electric polarization (P) and $2\times 2$ dyadic function (Q)}
\begin{align}
{\rm{Q}} = \eta \bigg(|cos\theta |\frac{{{{\rm{\textbf{k}}}_{\rm{t}}}{{\rm{\textbf{k}}}_{\rm{t}}}}}{{k_{\rm{t}}^2}} + \frac{1}{{|cos\theta |}}\frac{{{\rm{\hat {\textbf{n}} \times }}{{\rm{\textbf{k}}}_{\rm{t}}}{\rm{\hat {\textbf{n}} \times }}{{\rm{\textbf{k}}}_{\rm{t}}}}}{{k_{\rm{t}}^2}}\bigg)\ 
\end{align}
\begin{align}
{{\rm{\textbf{P}}}_{\rm{t}}} = \bigg[\hat \alpha _{{\rm{xx}}}^{{\rm{ee}}}\cos \theta  + \hat \alpha _{{\rm{xy}}}^{{\rm{ee}}}\sin \theta \bigg]\frac{{{{\rm{{E}}}^{\rm{i}}}}}{{\rm{S}}}{\rm{\hat {\textbf{x}}}}\
\end{align}
\begin{align}
{{\rm{P}}_{\rm{n}}} = \bigg[\hat \alpha _{{\rm{yx}}}^{{\rm{ee}}}\cos \theta  + \hat \alpha _{{\rm{yy}}}^{{\rm{ee}}}\sin \theta \bigg]\frac{{{{\rm{E}}^{\rm{i}}}}}{{\rm{S}}},\
\end{align}

\label{sec:refs}



\bibliography{sample}

\begin{thebibliography}{64}%
\makeatletter
\providecommand \@ifxundefined [1]{%
 \@ifx{#1\undefined}
}%
\providecommand \@ifnum [1]{%
 \ifnum #1\expandafter \@firstoftwo
 \else \expandafter \@secondoftwo
 \fi
}%
\providecommand \@ifx [1]{%
 \ifx #1\expandafter \@firstoftwo
 \else \expandafter \@secondoftwo
 \fi
}%
\providecommand \natexlab [1]{#1}%
\providecommand \enquote  [1]{``#1''}%
\providecommand \bibnamefont  [1]{#1}%
\providecommand \bibfnamefont [1]{#1}%
\providecommand \citenamefont [1]{#1}%
\providecommand \href@noop [0]{\@secondoftwo}%
\providecommand \href [0]{\begingroup \@sanitize@url \@href}%
\providecommand \@href[1]{\@@startlink{#1}\@@href}%
\providecommand \@@href[1]{\endgroup#1\@@endlink}%
\providecommand \@sanitize@url [0]{\catcode `\\12\catcode `\$12\catcode
  `\&12\catcode `\#12\catcode `\^12\catcode `\_12\catcode `\%12\relax}%
\providecommand \@@startlink[1]{}%
\providecommand \@@endlink[0]{}%
\providecommand \url  [0]{\begingroup\@sanitize@url \@url }%
\providecommand \@url [1]{\endgroup\@href {#1}{\urlprefix }}%
\providecommand \urlprefix  [0]{URL }%
\providecommand \Eprint [0]{\href }%
\providecommand \doibase [0]{http://dx.doi.org/}%
\providecommand \selectlanguage [0]{\@gobble}%
\providecommand \bibinfo  [0]{\@secondoftwo}%
\providecommand \bibfield  [0]{\@secondoftwo}%
\providecommand \translation [1]{[#1]}%
\providecommand \BibitemOpen [0]{}%
\providecommand \bibitemStop [0]{}%
\providecommand \bibitemNoStop [0]{.\EOS\space}%
\providecommand \EOS [0]{\spacefactor3000\relax}%
\providecommand \BibitemShut  [1]{\csname bibitem#1\endcsname}%
\let\auto@bib@innerbib\@empty
\bibitem [{\citenamefont {Balanis}(1992)}]{balanis}%
  \BibitemOpen
  \bibfield  {author} {\bibinfo {author} {\bibfnamefont {C.~A.}\ \bibnamefont
  {Balanis}},\ }\href@noop {} {\bibfield  {journal} {\bibinfo  {journal}
  {Proceedings of the IEEE}\ }\textbf {\bibinfo {volume} {80}},\ \bibinfo
  {pages} {7} (\bibinfo {year} {1992})}\BibitemShut {NoStop}%
\bibitem [{\citenamefont {Rotman}(1962)}]{rotman}%
  \BibitemOpen
  \bibfield  {author} {\bibinfo {author} {\bibfnamefont {W.}~\bibnamefont
  {Rotman}},\ }\href@noop {} {\bibfield  {journal} {\bibinfo  {journal} {IRE
  Transactions on Antennas and Propagation}\ }\textbf {\bibinfo {volume}
  {10}},\ \bibinfo {pages} {82} (\bibinfo {year} {1962})}\BibitemShut {NoStop}%
\bibitem [{\citenamefont {Smith}\ and\ \citenamefont {Schurig}(2003)}]{Smith}%
  \BibitemOpen
  \bibfield  {author} {\bibinfo {author} {\bibfnamefont {D.}~\bibnamefont
  {Smith}}\ and\ \bibinfo {author} {\bibfnamefont {D.}~\bibnamefont
  {Schurig}},\ }\href@noop {} {\bibfield  {journal} {\bibinfo  {journal}
  {Physical Review Letters}\ }\textbf {\bibinfo {volume} {90}},\ \bibinfo
  {pages} {077405} (\bibinfo {year} {2003})}\BibitemShut {NoStop}%
\bibitem [{\citenamefont {Lu}\ and\ \citenamefont {Sridhar}(2008)}]{Lu}%
  \BibitemOpen
  \bibfield  {author} {\bibinfo {author} {\bibfnamefont {W.}~\bibnamefont
  {Lu}}\ and\ \bibinfo {author} {\bibfnamefont {S.}~\bibnamefont {Sridhar}},\
  }\href@noop {} {\bibfield  {journal} {\bibinfo  {journal} {Physical Review
  B}\ }\textbf {\bibinfo {volume} {77}},\ \bibinfo {pages} {233101} (\bibinfo
  {year} {2008})}\BibitemShut {NoStop}%
\bibitem [{\citenamefont {Poddubny}\ \emph {et~al.}(2013)\citenamefont
  {Poddubny}, \citenamefont {Iorsh}, \citenamefont {Belov},\ and\ \citenamefont
  {Kivshar}}]{Poddubny}%
  \BibitemOpen
  \bibfield  {author} {\bibinfo {author} {\bibfnamefont {A.}~\bibnamefont
  {Poddubny}}, \bibinfo {author} {\bibfnamefont {I.}~\bibnamefont {Iorsh}},
  \bibinfo {author} {\bibfnamefont {P.}~\bibnamefont {Belov}}, \ and\ \bibinfo
  {author} {\bibfnamefont {Y.}~\bibnamefont {Kivshar}},\ }\href@noop {}
  {\bibfield  {journal} {\bibinfo  {journal} {Nature photonics}\ }\textbf
  {\bibinfo {volume} {7}},\ \bibinfo {pages} {948} (\bibinfo {year}
  {2013})}\BibitemShut {NoStop}%
\bibitem [{\citenamefont {Simovski}\ \emph {et~al.}(2013)\citenamefont
  {Simovski}, \citenamefont {Maslovski}, \citenamefont {Nefedov},\ and\
  \citenamefont {Tretyakov}}]{Simovski}%
  \BibitemOpen
  \bibfield  {author} {\bibinfo {author} {\bibfnamefont {C.}~\bibnamefont
  {Simovski}}, \bibinfo {author} {\bibfnamefont {S.}~\bibnamefont {Maslovski}},
  \bibinfo {author} {\bibfnamefont {I.}~\bibnamefont {Nefedov}}, \ and\
  \bibinfo {author} {\bibfnamefont {S.}~\bibnamefont {Tretyakov}},\ }\href@noop
  {} {\bibfield  {journal} {\bibinfo  {journal} {Optics express}\ }\textbf
  {\bibinfo {volume} {21}},\ \bibinfo {pages} {14988} (\bibinfo {year}
  {2013})}\BibitemShut {NoStop}%
\bibitem [{\citenamefont {Morgado}\ \emph {et~al.}(2011)\citenamefont
  {Morgado}, \citenamefont {Marcos}, \citenamefont {Silveirinha},\ and\
  \citenamefont {Maslovski}}]{Morgado}%
  \BibitemOpen
  \bibfield  {author} {\bibinfo {author} {\bibfnamefont {T.~A.}\ \bibnamefont
  {Morgado}}, \bibinfo {author} {\bibfnamefont {J.~S.}\ \bibnamefont {Marcos}},
  \bibinfo {author} {\bibfnamefont {M.~G.}\ \bibnamefont {Silveirinha}}, \ and\
  \bibinfo {author} {\bibfnamefont {S.~I.}\ \bibnamefont {Maslovski}},\
  }\href@noop {} {\bibfield  {journal} {\bibinfo  {journal} {Physical review
  letters}\ }\textbf {\bibinfo {volume} {107}},\ \bibinfo {pages} {063903}
  (\bibinfo {year} {2011})}\BibitemShut {NoStop}%
\bibitem [{\citenamefont {Pendry}\ \emph {et~al.}(1996)\citenamefont {Pendry},
  \citenamefont {Holden}, \citenamefont {Stewart},\ and\ \citenamefont
  {Youngs}}]{Pendry}%
  \BibitemOpen
  \bibfield  {author} {\bibinfo {author} {\bibfnamefont {J.~B.}\ \bibnamefont
  {Pendry}}, \bibinfo {author} {\bibfnamefont {A.}~\bibnamefont {Holden}},
  \bibinfo {author} {\bibfnamefont {W.}~\bibnamefont {Stewart}}, \ and\
  \bibinfo {author} {\bibfnamefont {I.}~\bibnamefont {Youngs}},\ }\href@noop {}
  {\bibfield  {journal} {\bibinfo  {journal} {Physical review letters}\
  }\textbf {\bibinfo {volume} {76}},\ \bibinfo {pages} {4773} (\bibinfo {year}
  {1996})}\BibitemShut {NoStop}%
\bibitem [{\citenamefont {Silveirinha}\ and\ \citenamefont
  {Fernandes}(2008)}]{Silveirinha}%
  \BibitemOpen
  \bibfield  {author} {\bibinfo {author} {\bibfnamefont {M.~G.}\ \bibnamefont
  {Silveirinha}}\ and\ \bibinfo {author} {\bibfnamefont {C.~A.}\ \bibnamefont
  {Fernandes}},\ }\href@noop {} {\bibfield  {journal} {\bibinfo  {journal}
  {Physical Review B}\ }\textbf {\bibinfo {volume} {78}},\ \bibinfo {pages}
  {033108} (\bibinfo {year} {2008})}\BibitemShut {NoStop}%
\bibitem [{\citenamefont {Laroche}\ \emph
  {et~al.}(2006{\natexlab{a}})\citenamefont {Laroche}, \citenamefont
  {Albaladejo}, \citenamefont {G{\'o}mez-Medina},\ and\ \citenamefont
  {S{\'a}enz}}]{Laroche}%
  \BibitemOpen
  \bibfield  {author} {\bibinfo {author} {\bibfnamefont {M.}~\bibnamefont
  {Laroche}}, \bibinfo {author} {\bibfnamefont {S.}~\bibnamefont {Albaladejo}},
  \bibinfo {author} {\bibfnamefont {R.}~\bibnamefont {G{\'o}mez-Medina}}, \
  and\ \bibinfo {author} {\bibfnamefont {J.~J.}\ \bibnamefont {S{\'a}enz}},\
  }\href@noop {} {\bibfield  {journal} {\bibinfo  {journal} {Physical Review
  B}\ }\textbf {\bibinfo {volume} {74}},\ \bibinfo {pages} {245422} (\bibinfo
  {year} {2006}{\natexlab{a}})}\BibitemShut {NoStop}%
\bibitem [{\citenamefont {Ghenuche}\ \emph {et~al.}(2012)\citenamefont
  {Ghenuche}, \citenamefont {Vincent}, \citenamefont {Laroche}, \citenamefont
  {Bardou}, \citenamefont {Ha{\"\i}dar}, \citenamefont {Pelouard},\ and\
  \citenamefont {Collin}}]{Ghenuche}%
  \BibitemOpen
  \bibfield  {author} {\bibinfo {author} {\bibfnamefont {P.}~\bibnamefont
  {Ghenuche}}, \bibinfo {author} {\bibfnamefont {G.}~\bibnamefont {Vincent}},
  \bibinfo {author} {\bibfnamefont {M.}~\bibnamefont {Laroche}}, \bibinfo
  {author} {\bibfnamefont {N.}~\bibnamefont {Bardou}}, \bibinfo {author}
  {\bibfnamefont {R.}~\bibnamefont {Ha{\"\i}dar}}, \bibinfo {author}
  {\bibfnamefont {J.-L.}\ \bibnamefont {Pelouard}}, \ and\ \bibinfo {author}
  {\bibfnamefont {S.}~\bibnamefont {Collin}},\ }\href@noop {} {\bibfield
  {journal} {\bibinfo  {journal} {Physical review letters}\ }\textbf {\bibinfo
  {volume} {109}},\ \bibinfo {pages} {143903} (\bibinfo {year}
  {2012})}\BibitemShut {NoStop}%
\bibitem [{\citenamefont {McMahon}\ \emph {et~al.}(2009)\citenamefont
  {McMahon}, \citenamefont {Gray},\ and\ \citenamefont {Schatz}}]{McMahon}%
  \BibitemOpen
  \bibfield  {author} {\bibinfo {author} {\bibfnamefont {J.~M.}\ \bibnamefont
  {McMahon}}, \bibinfo {author} {\bibfnamefont {S.~K.}\ \bibnamefont {Gray}}, \
  and\ \bibinfo {author} {\bibfnamefont {G.~C.}\ \bibnamefont {Schatz}},\
  }\href@noop {} {\bibfield  {journal} {\bibinfo  {journal} {Physical review
  letters}\ }\textbf {\bibinfo {volume} {103}},\ \bibinfo {pages} {097403}
  (\bibinfo {year} {2009})}\BibitemShut {NoStop}%
\bibitem [{\citenamefont {Ikonen}\ \emph {et~al.}(2004)\citenamefont {Ikonen},
  \citenamefont {Simovski},\ and\ \citenamefont {Tretyakov}}]{Ikonen}%
  \BibitemOpen
  \bibfield  {author} {\bibinfo {author} {\bibfnamefont {P.}~\bibnamefont
  {Ikonen}}, \bibinfo {author} {\bibfnamefont {C.}~\bibnamefont {Simovski}}, \
  and\ \bibinfo {author} {\bibfnamefont {S.}~\bibnamefont {Tretyakov}},\
  }\href@noop {} {\bibfield  {journal} {\bibinfo  {journal} {Microwave and
  Optical Technology Letters}\ }\textbf {\bibinfo {volume} {43}},\ \bibinfo
  {pages} {467} (\bibinfo {year} {2004})}\BibitemShut {NoStop}%
\bibitem [{\citenamefont {Capolino}(2017)}]{Capolino}%
  \BibitemOpen
  \bibfield  {author} {\bibinfo {author} {\bibfnamefont {F.}~\bibnamefont
  {Capolino}},\ }\href@noop {} {\emph {\bibinfo {title} {Theory and phenomena
  of metamaterials}}}\ (\bibinfo  {publisher} {CRC press},\ \bibinfo {year}
  {2017})\BibitemShut {NoStop}%
\bibitem [{\citenamefont {Salary}\ and\ \citenamefont
  {Mosallaei}(2017)}]{Salary}%
  \BibitemOpen
  \bibfield  {author} {\bibinfo {author} {\bibfnamefont {M.~M.}\ \bibnamefont
  {Salary}}\ and\ \bibinfo {author} {\bibfnamefont {H.}~\bibnamefont
  {Mosallaei}},\ }\href@noop {} {\bibfield  {journal} {\bibinfo  {journal}
  {Scientific reports}\ }\textbf {\bibinfo {volume} {7}},\ \bibinfo {pages}
  {10055} (\bibinfo {year} {2017})}\BibitemShut {NoStop}%
\bibitem [{\citenamefont {Momeni}\ \emph {et~al.}(2018)\citenamefont {Momeni},
  \citenamefont {Rouhi}, \citenamefont {Rajabalipanah},\ and\ \citenamefont
  {Abdolali}}]{momeni}%
  \BibitemOpen
  \bibfield  {author} {\bibinfo {author} {\bibfnamefont {A.}~\bibnamefont
  {Momeni}}, \bibinfo {author} {\bibfnamefont {K.}~\bibnamefont {Rouhi}},
  \bibinfo {author} {\bibfnamefont {H.}~\bibnamefont {Rajabalipanah}}, \ and\
  \bibinfo {author} {\bibfnamefont {A.}~\bibnamefont {Abdolali}},\ }\href@noop
  {} {\bibfield  {journal} {\bibinfo  {journal} {Scientific reports}\ }\textbf
  {\bibinfo {volume} {8}},\ \bibinfo {pages} {6200} (\bibinfo {year}
  {2018})}\BibitemShut {NoStop}%
\bibitem [{\citenamefont {Zhang}\ \emph {et~al.}(2013)\citenamefont {Zhang},
  \citenamefont {MacDonald},\ and\ \citenamefont {Zheludev}}]{Zhang}%
  \BibitemOpen
  \bibfield  {author} {\bibinfo {author} {\bibfnamefont {J.}~\bibnamefont
  {Zhang}}, \bibinfo {author} {\bibfnamefont {K.~F.}\ \bibnamefont
  {MacDonald}}, \ and\ \bibinfo {author} {\bibfnamefont {N.~I.}\ \bibnamefont
  {Zheludev}},\ }\href@noop {} {\bibfield  {journal} {\bibinfo  {journal}
  {Optics express}\ }\textbf {\bibinfo {volume} {21}},\ \bibinfo {pages}
  {26721} (\bibinfo {year} {2013})}\BibitemShut {NoStop}%
\bibitem [{\citenamefont {Pors}\ and\ \citenamefont
  {Bozhevolnyi}(2013)}]{Pors}%
  \BibitemOpen
  \bibfield  {author} {\bibinfo {author} {\bibfnamefont {A.}~\bibnamefont
  {Pors}}\ and\ \bibinfo {author} {\bibfnamefont {S.~I.}\ \bibnamefont
  {Bozhevolnyi}},\ }\href@noop {} {\bibfield  {journal} {\bibinfo  {journal}
  {Optics express}\ }\textbf {\bibinfo {volume} {21}},\ \bibinfo {pages}
  {27438} (\bibinfo {year} {2013})}\BibitemShut {NoStop}%
\bibitem [{\citenamefont {Du}\ \emph {et~al.}(2013)\citenamefont {Du},
  \citenamefont {Lin}, \citenamefont {Chui}, \citenamefont {Dong},\ and\
  \citenamefont {Zhang}}]{Lin}%
  \BibitemOpen
  \bibfield  {author} {\bibinfo {author} {\bibfnamefont {J.}~\bibnamefont
  {Du}}, \bibinfo {author} {\bibfnamefont {Z.}~\bibnamefont {Lin}}, \bibinfo
  {author} {\bibfnamefont {S.}~\bibnamefont {Chui}}, \bibinfo {author}
  {\bibfnamefont {G.}~\bibnamefont {Dong}}, \ and\ \bibinfo {author}
  {\bibfnamefont {W.}~\bibnamefont {Zhang}},\ }\href@noop {} {\bibfield
  {journal} {\bibinfo  {journal} {Physical review letters}\ }\textbf {\bibinfo
  {volume} {110}},\ \bibinfo {pages} {163902} (\bibinfo {year}
  {2013})}\BibitemShut {NoStop}%
\bibitem [{\citenamefont {Yanik}\ \emph {et~al.}(2009)\citenamefont {Yanik},
  \citenamefont {Adato}, \citenamefont {Erramilli},\ and\ \citenamefont
  {Altug}}]{Yanik}%
  \BibitemOpen
  \bibfield  {author} {\bibinfo {author} {\bibfnamefont {A.~A.}\ \bibnamefont
  {Yanik}}, \bibinfo {author} {\bibfnamefont {R.}~\bibnamefont {Adato}},
  \bibinfo {author} {\bibfnamefont {S.}~\bibnamefont {Erramilli}}, \ and\
  \bibinfo {author} {\bibfnamefont {H.}~\bibnamefont {Altug}},\ }\href@noop {}
  {\bibfield  {journal} {\bibinfo  {journal} {Optics express}\ }\textbf
  {\bibinfo {volume} {17}},\ \bibinfo {pages} {20900} (\bibinfo {year}
  {2009})}\BibitemShut {NoStop}%
\bibitem [{\citenamefont {Rajabalipanah}\ \emph {et~al.}(2019)\citenamefont
  {Rajabalipanah}, \citenamefont {Abdolali}, \citenamefont {Shabanpour},
  \citenamefont {Momeni},\ and\ \citenamefont {Cheldavi}}]{Rajabalipanah}%
  \BibitemOpen
  \bibfield  {author} {\bibinfo {author} {\bibfnamefont {H.}~\bibnamefont
  {Rajabalipanah}}, \bibinfo {author} {\bibfnamefont {A.}~\bibnamefont
  {Abdolali}}, \bibinfo {author} {\bibfnamefont {J.}~\bibnamefont
  {Shabanpour}}, \bibinfo {author} {\bibfnamefont {A.}~\bibnamefont {Momeni}},
  \ and\ \bibinfo {author} {\bibfnamefont {A.}~\bibnamefont {Cheldavi}},\
  }\href@noop {} {\bibfield  {journal} {\bibinfo  {journal} {ACS omega}\ }
  (\bibinfo {year} {2019})}\BibitemShut {NoStop}%
\bibitem [{\citenamefont {Casse}\ \emph {et~al.}(2010)\citenamefont {Casse},
  \citenamefont {Lu}, \citenamefont {Huang}, \citenamefont {Gultepe},
  \citenamefont {Menon},\ and\ \citenamefont {Sridhar}}]{Casse}%
  \BibitemOpen
  \bibfield  {author} {\bibinfo {author} {\bibfnamefont {B.}~\bibnamefont
  {Casse}}, \bibinfo {author} {\bibfnamefont {W.}~\bibnamefont {Lu}}, \bibinfo
  {author} {\bibfnamefont {Y.}~\bibnamefont {Huang}}, \bibinfo {author}
  {\bibfnamefont {E.}~\bibnamefont {Gultepe}}, \bibinfo {author} {\bibfnamefont
  {L.}~\bibnamefont {Menon}}, \ and\ \bibinfo {author} {\bibfnamefont
  {S.}~\bibnamefont {Sridhar}},\ }\href@noop {} {\bibfield  {journal} {\bibinfo
   {journal} {Applied Physics Letters}\ }\textbf {\bibinfo {volume} {96}},\
  \bibinfo {pages} {023114} (\bibinfo {year} {2010})}\BibitemShut {NoStop}%
\bibitem [{\citenamefont {Shvets}\ \emph {et~al.}(2007)\citenamefont {Shvets},
  \citenamefont {Trendafilov}, \citenamefont {Pendry},\ and\ \citenamefont
  {Sarychev}}]{Shvets}%
  \BibitemOpen
  \bibfield  {author} {\bibinfo {author} {\bibfnamefont {G.}~\bibnamefont
  {Shvets}}, \bibinfo {author} {\bibfnamefont {S.}~\bibnamefont {Trendafilov}},
  \bibinfo {author} {\bibfnamefont {J.}~\bibnamefont {Pendry}}, \ and\ \bibinfo
  {author} {\bibfnamefont {A.}~\bibnamefont {Sarychev}},\ }\href@noop {}
  {\bibfield  {journal} {\bibinfo  {journal} {Physical review letters}\
  }\textbf {\bibinfo {volume} {99}},\ \bibinfo {pages} {053903} (\bibinfo
  {year} {2007})}\BibitemShut {NoStop}%
\bibitem [{\citenamefont {Silveirinha}(2006)}]{Silveirinha1}%
  \BibitemOpen
  \bibfield  {author} {\bibinfo {author} {\bibfnamefont {M.~G.}\ \bibnamefont
  {Silveirinha}},\ }\href@noop {} {\bibfield  {journal} {\bibinfo  {journal}
  {Physical Review E}\ }\textbf {\bibinfo {volume} {73}},\ \bibinfo {pages}
  {046612} (\bibinfo {year} {2006})}\BibitemShut {NoStop}%
\bibitem [{\citenamefont {Maslovski}\ and\ \citenamefont
  {Silveirinha}(2011)}]{Maslovski}%
  \BibitemOpen
  \bibfield  {author} {\bibinfo {author} {\bibfnamefont {S.~I.}\ \bibnamefont
  {Maslovski}}\ and\ \bibinfo {author} {\bibfnamefont {M.~G.}\ \bibnamefont
  {Silveirinha}},\ }\href@noop {} {\bibfield  {journal} {\bibinfo  {journal}
  {Physical Review A}\ }\textbf {\bibinfo {volume} {83}},\ \bibinfo {pages}
  {022508} (\bibinfo {year} {2011})}\BibitemShut {NoStop}%
\bibitem [{\citenamefont {Yao}\ \emph {et~al.}(2011)\citenamefont {Yao},
  \citenamefont {Yang}, \citenamefont {Yin}, \citenamefont {Bartal},\ and\
  \citenamefont {Zhang}}]{Yao}%
  \BibitemOpen
  \bibfield  {author} {\bibinfo {author} {\bibfnamefont {J.}~\bibnamefont
  {Yao}}, \bibinfo {author} {\bibfnamefont {X.}~\bibnamefont {Yang}}, \bibinfo
  {author} {\bibfnamefont {X.}~\bibnamefont {Yin}}, \bibinfo {author}
  {\bibfnamefont {G.}~\bibnamefont {Bartal}}, \ and\ \bibinfo {author}
  {\bibfnamefont {X.}~\bibnamefont {Zhang}},\ }\href@noop {} {\bibfield
  {journal} {\bibinfo  {journal} {Proceedings of the National Academy of
  Sciences}\ }\textbf {\bibinfo {volume} {108}},\ \bibinfo {pages} {11327}
  (\bibinfo {year} {2011})}\BibitemShut {NoStop}%
\bibitem [{\citenamefont {Jacob}\ \emph {et~al.}(2012)\citenamefont {Jacob},
  \citenamefont {Smolyaninov},\ and\ \citenamefont {Narimanov}}]{Jacob}%
  \BibitemOpen
  \bibfield  {author} {\bibinfo {author} {\bibfnamefont {Z.}~\bibnamefont
  {Jacob}}, \bibinfo {author} {\bibfnamefont {I.~I.}\ \bibnamefont
  {Smolyaninov}}, \ and\ \bibinfo {author} {\bibfnamefont {E.~E.}\ \bibnamefont
  {Narimanov}},\ }\href@noop {} {\bibfield  {journal} {\bibinfo  {journal}
  {Applied Physics Letters}\ }\textbf {\bibinfo {volume} {100}},\ \bibinfo
  {pages} {181105} (\bibinfo {year} {2012})}\BibitemShut {NoStop}%
\bibitem [{\citenamefont {Cortes}\ \emph {et~al.}(2012)\citenamefont {Cortes},
  \citenamefont {Newman}, \citenamefont {Molesky},\ and\ \citenamefont
  {Jacob}}]{Cortes}%
  \BibitemOpen
  \bibfield  {author} {\bibinfo {author} {\bibfnamefont {C.}~\bibnamefont
  {Cortes}}, \bibinfo {author} {\bibfnamefont {W.}~\bibnamefont {Newman}},
  \bibinfo {author} {\bibfnamefont {S.}~\bibnamefont {Molesky}}, \ and\
  \bibinfo {author} {\bibfnamefont {Z.}~\bibnamefont {Jacob}},\ }\href@noop {}
  {\bibfield  {journal} {\bibinfo  {journal} {Journal of Optics}\ }\textbf
  {\bibinfo {volume} {14}},\ \bibinfo {pages} {063001} (\bibinfo {year}
  {2012})}\BibitemShut {NoStop}%
\bibitem [{\citenamefont {Kabashin}\ \emph {et~al.}(2009)\citenamefont
  {Kabashin}, \citenamefont {Evans}, \citenamefont {Pastkovsky}, \citenamefont
  {Hendren}, \citenamefont {Wurtz}, \citenamefont {Atkinson}, \citenamefont
  {Pollard}, \citenamefont {Podolskiy},\ and\ \citenamefont
  {Zayats}}]{Kabashin}%
  \BibitemOpen
  \bibfield  {author} {\bibinfo {author} {\bibfnamefont {A.}~\bibnamefont
  {Kabashin}}, \bibinfo {author} {\bibfnamefont {P.}~\bibnamefont {Evans}},
  \bibinfo {author} {\bibfnamefont {S.}~\bibnamefont {Pastkovsky}}, \bibinfo
  {author} {\bibfnamefont {W.}~\bibnamefont {Hendren}}, \bibinfo {author}
  {\bibfnamefont {G.}~\bibnamefont {Wurtz}}, \bibinfo {author} {\bibfnamefont
  {R.}~\bibnamefont {Atkinson}}, \bibinfo {author} {\bibfnamefont
  {R.}~\bibnamefont {Pollard}}, \bibinfo {author} {\bibfnamefont
  {V.}~\bibnamefont {Podolskiy}}, \ and\ \bibinfo {author} {\bibfnamefont
  {A.}~\bibnamefont {Zayats}},\ }\href@noop {} {\bibfield  {journal} {\bibinfo
  {journal} {Nature materials}\ }\textbf {\bibinfo {volume} {8}},\ \bibinfo
  {pages} {867} (\bibinfo {year} {2009})}\BibitemShut {NoStop}%
\bibitem [{\citenamefont {Kallos}\ \emph {et~al.}(2012)\citenamefont {Kallos},
  \citenamefont {Chremmos},\ and\ \citenamefont {Yannopapas}}]{Kallos}%
  \BibitemOpen
  \bibfield  {author} {\bibinfo {author} {\bibfnamefont {E.}~\bibnamefont
  {Kallos}}, \bibinfo {author} {\bibfnamefont {I.}~\bibnamefont {Chremmos}}, \
  and\ \bibinfo {author} {\bibfnamefont {V.}~\bibnamefont {Yannopapas}},\
  }\href@noop {} {\bibfield  {journal} {\bibinfo  {journal} {Physical Review
  B}\ }\textbf {\bibinfo {volume} {86}},\ \bibinfo {pages} {245108} (\bibinfo
  {year} {2012})}\BibitemShut {NoStop}%
\bibitem [{\citenamefont {Strickland}\ \emph {et~al.}(2015)\citenamefont
  {Strickland}, \citenamefont {Ay{\'o}n},\ and\ \citenamefont
  {Al{\`u}}}]{Strickland}%
  \BibitemOpen
  \bibfield  {author} {\bibinfo {author} {\bibfnamefont {D.}~\bibnamefont
  {Strickland}}, \bibinfo {author} {\bibfnamefont {A.}~\bibnamefont
  {Ay{\'o}n}}, \ and\ \bibinfo {author} {\bibfnamefont {A.}~\bibnamefont
  {Al{\`u}}},\ }\href@noop {} {\bibfield  {journal} {\bibinfo  {journal}
  {Physical Review B}\ }\textbf {\bibinfo {volume} {91}},\ \bibinfo {pages}
  {085104} (\bibinfo {year} {2015})}\BibitemShut {NoStop}%
\bibitem [{\citenamefont {Raab}\ \emph {et~al.}(2005)\citenamefont {Raab},
  \citenamefont {De~Lange},\ and\ \citenamefont {de~Lange}}]{Raab}%
  \BibitemOpen
  \bibfield  {author} {\bibinfo {author} {\bibfnamefont {R.~E.}\ \bibnamefont
  {Raab}}, \bibinfo {author} {\bibfnamefont {O.~L.}\ \bibnamefont {De~Lange}},
  \ and\ \bibinfo {author} {\bibfnamefont {O.~L.}\ \bibnamefont {de~Lange}},\
  }\href@noop {} {\emph {\bibinfo {title} {Multipole theory in
  electromagnetism: classical, quantum, and symmetry aspects, with
  applications}}},\ Vol.\ \bibinfo {volume} {128}\ (\bibinfo  {publisher}
  {Oxford University Press on Demand},\ \bibinfo {year} {2005})\BibitemShut
  {NoStop}%
\bibitem [{\citenamefont {Barron}(2009)}]{Barron}%
  \BibitemOpen
  \bibfield  {author} {\bibinfo {author} {\bibfnamefont {L.~D.}\ \bibnamefont
  {Barron}},\ }\href@noop {} {\emph {\bibinfo {title} {Molecular light
  scattering and optical activity}}}\ (\bibinfo  {publisher} {Cambridge
  University Press},\ \bibinfo {year} {2009})\BibitemShut {NoStop}%
\bibitem [{\citenamefont {Silva}\ \emph {et~al.}(2014)\citenamefont {Silva},
  \citenamefont {Monticone}, \citenamefont {Castaldi}, \citenamefont {Galdi},
  \citenamefont {Al{\`u}},\ and\ \citenamefont
  {Engheta}}]{silva2014performing}%
  \BibitemOpen
  \bibfield  {author} {\bibinfo {author} {\bibfnamefont {A.}~\bibnamefont
  {Silva}}, \bibinfo {author} {\bibfnamefont {F.}~\bibnamefont {Monticone}},
  \bibinfo {author} {\bibfnamefont {G.}~\bibnamefont {Castaldi}}, \bibinfo
  {author} {\bibfnamefont {V.}~\bibnamefont {Galdi}}, \bibinfo {author}
  {\bibfnamefont {A.}~\bibnamefont {Al{\`u}}}, \ and\ \bibinfo {author}
  {\bibfnamefont {N.}~\bibnamefont {Engheta}},\ }\href@noop {} {\bibfield
  {journal} {\bibinfo  {journal} {Science}\ }\textbf {\bibinfo {volume}
  {343}},\ \bibinfo {pages} {160} (\bibinfo {year} {2014})}\BibitemShut
  {NoStop}%
\bibitem [{\citenamefont {Pors}\ \emph {et~al.}(2015)\citenamefont {Pors},
  \citenamefont {Nielsen},\ and\ \citenamefont {Bozhevolnyi}}]{pors2015analog}%
  \BibitemOpen
  \bibfield  {author} {\bibinfo {author} {\bibfnamefont {A.}~\bibnamefont
  {Pors}}, \bibinfo {author} {\bibfnamefont {M.~G.}\ \bibnamefont {Nielsen}}, \
  and\ \bibinfo {author} {\bibfnamefont {S.~I.}\ \bibnamefont {Bozhevolnyi}},\
  }\href@noop {} {\bibfield  {journal} {\bibinfo  {journal} {Nano letters}\
  }\textbf {\bibinfo {volume} {15}},\ \bibinfo {pages} {791} (\bibinfo {year}
  {2015})}\BibitemShut {NoStop}%
\bibitem [{\citenamefont {Babaee}\ \emph {et~al.}(2020)\citenamefont {Babaee},
  \citenamefont {Momeni}, \citenamefont {Abdolali},\ and\ \citenamefont
  {Fleury}}]{babaee2020parallel}%
  \BibitemOpen
  \bibfield  {author} {\bibinfo {author} {\bibfnamefont {A.}~\bibnamefont
  {Babaee}}, \bibinfo {author} {\bibfnamefont {A.}~\bibnamefont {Momeni}},
  \bibinfo {author} {\bibfnamefont {A.}~\bibnamefont {Abdolali}}, \ and\
  \bibinfo {author} {\bibfnamefont {R.}~\bibnamefont {Fleury}},\ }\href@noop {}
  {\bibfield  {journal} {\bibinfo  {journal} {arXiv preprint arXiv:2004.02948}\
  } (\bibinfo {year} {2020})}\BibitemShut {NoStop}%
\bibitem [{\citenamefont {Kwon}\ \emph {et~al.}(2018)\citenamefont {Kwon},
  \citenamefont {Sounas}, \citenamefont {Cordaro}, \citenamefont {Polman},\
  and\ \citenamefont {Al{\`u}}}]{kwon2018nonlocal}%
  \BibitemOpen
  \bibfield  {author} {\bibinfo {author} {\bibfnamefont {H.}~\bibnamefont
  {Kwon}}, \bibinfo {author} {\bibfnamefont {D.}~\bibnamefont {Sounas}},
  \bibinfo {author} {\bibfnamefont {A.}~\bibnamefont {Cordaro}}, \bibinfo
  {author} {\bibfnamefont {A.}~\bibnamefont {Polman}}, \ and\ \bibinfo {author}
  {\bibfnamefont {A.}~\bibnamefont {Al{\`u}}},\ }\href@noop {} {\bibfield
  {journal} {\bibinfo  {journal} {Physical review letters}\ }\textbf {\bibinfo
  {volume} {121}},\ \bibinfo {pages} {173004} (\bibinfo {year}
  {2018})}\BibitemShut {NoStop}%
\bibitem [{\citenamefont {Zhu}\ \emph {et~al.}(2017)\citenamefont {Zhu},
  \citenamefont {Zhou}, \citenamefont {Lou}, \citenamefont {Ye}, \citenamefont
  {Qiu}, \citenamefont {Ruan},\ and\ \citenamefont {Fan}}]{zhu2017plasmonic}%
  \BibitemOpen
  \bibfield  {author} {\bibinfo {author} {\bibfnamefont {T.}~\bibnamefont
  {Zhu}}, \bibinfo {author} {\bibfnamefont {Y.}~\bibnamefont {Zhou}}, \bibinfo
  {author} {\bibfnamefont {Y.}~\bibnamefont {Lou}}, \bibinfo {author}
  {\bibfnamefont {H.}~\bibnamefont {Ye}}, \bibinfo {author} {\bibfnamefont
  {M.}~\bibnamefont {Qiu}}, \bibinfo {author} {\bibfnamefont {Z.}~\bibnamefont
  {Ruan}}, \ and\ \bibinfo {author} {\bibfnamefont {S.}~\bibnamefont {Fan}},\
  }\href@noop {} {\bibfield  {journal} {\bibinfo  {journal} {Nature
  communications}\ }\textbf {\bibinfo {volume} {8}},\ \bibinfo {pages} {1}
  (\bibinfo {year} {2017})}\BibitemShut {NoStop}%
\bibitem [{\citenamefont {Momeni}\ \emph {et~al.}(2019)\citenamefont {Momeni},
  \citenamefont {Rajabalipanah}, \citenamefont {Abdolali},\ and\ \citenamefont
  {Achouri}}]{momeni2}%
  \BibitemOpen
  \bibfield  {author} {\bibinfo {author} {\bibfnamefont {A.}~\bibnamefont
  {Momeni}}, \bibinfo {author} {\bibfnamefont {H.}~\bibnamefont
  {Rajabalipanah}}, \bibinfo {author} {\bibfnamefont {A.}~\bibnamefont
  {Abdolali}}, \ and\ \bibinfo {author} {\bibfnamefont {K.}~\bibnamefont
  {Achouri}},\ }\href@noop {} {\bibfield  {journal} {\bibinfo  {journal}
  {Physical Review Applied}\ }\textbf {\bibinfo {volume} {11}},\ \bibinfo
  {pages} {064042} (\bibinfo {year} {2019})}\BibitemShut {NoStop}%
\bibitem [{\citenamefont {Abdolali}\ \emph {et~al.}(2019)\citenamefont
  {Abdolali}, \citenamefont {Momeni}, \citenamefont {Rajabalipanah},\ and\
  \citenamefont {Achouri}}]{momeni3}%
  \BibitemOpen
  \bibfield  {author} {\bibinfo {author} {\bibfnamefont {A.}~\bibnamefont
  {Abdolali}}, \bibinfo {author} {\bibfnamefont {A.}~\bibnamefont {Momeni}},
  \bibinfo {author} {\bibfnamefont {H.}~\bibnamefont {Rajabalipanah}}, \ and\
  \bibinfo {author} {\bibfnamefont {K.}~\bibnamefont {Achouri}},\ }\href@noop
  {} {\bibfield  {journal} {\bibinfo  {journal} {New Journal of Physics}\ }
  (\bibinfo {year} {2019})}\BibitemShut {NoStop}%
\bibitem [{\citenamefont {Cordaro}\ \emph {et~al.}(2019)\citenamefont
  {Cordaro}, \citenamefont {Kwon}, \citenamefont {Sounas}, \citenamefont
  {Koenderink}, \citenamefont {Al{\`u}},\ and\ \citenamefont
  {Polman}}]{cordaro2019high}%
  \BibitemOpen
  \bibfield  {author} {\bibinfo {author} {\bibfnamefont {A.}~\bibnamefont
  {Cordaro}}, \bibinfo {author} {\bibfnamefont {H.}~\bibnamefont {Kwon}},
  \bibinfo {author} {\bibfnamefont {D.}~\bibnamefont {Sounas}}, \bibinfo
  {author} {\bibfnamefont {A.~F.}\ \bibnamefont {Koenderink}}, \bibinfo
  {author} {\bibfnamefont {A.}~\bibnamefont {Al{\`u}}}, \ and\ \bibinfo
  {author} {\bibfnamefont {A.}~\bibnamefont {Polman}},\ }\href@noop {}
  {\bibfield  {journal} {\bibinfo  {journal} {Nano letters}\ }\textbf {\bibinfo
  {volume} {19}},\ \bibinfo {pages} {8418} (\bibinfo {year}
  {2019})}\BibitemShut {NoStop}%
\bibitem [{\citenamefont {He}\ \emph {et~al.}(2020)\citenamefont {He},
  \citenamefont {Zhou}, \citenamefont {Chen}, \citenamefont {Shu},
  \citenamefont {Luo},\ and\ \citenamefont {Wen}}]{he2020spatial}%
  \BibitemOpen
  \bibfield  {author} {\bibinfo {author} {\bibfnamefont {S.}~\bibnamefont
  {He}}, \bibinfo {author} {\bibfnamefont {J.}~\bibnamefont {Zhou}}, \bibinfo
  {author} {\bibfnamefont {S.}~\bibnamefont {Chen}}, \bibinfo {author}
  {\bibfnamefont {W.}~\bibnamefont {Shu}}, \bibinfo {author} {\bibfnamefont
  {H.}~\bibnamefont {Luo}}, \ and\ \bibinfo {author} {\bibfnamefont
  {S.}~\bibnamefont {Wen}},\ }\href@noop {} {\bibfield  {journal} {\bibinfo
  {journal} {Optics Letters}\ }\textbf {\bibinfo {volume} {45}},\ \bibinfo
  {pages} {877} (\bibinfo {year} {2020})}\BibitemShut {NoStop}%
\bibitem [{\citenamefont {Zhu}\ \emph {et~al.}(2019)\citenamefont {Zhu},
  \citenamefont {Lou}, \citenamefont {Zhou}, \citenamefont {Zhang},
  \citenamefont {Huang}, \citenamefont {Li}, \citenamefont {Luo}, \citenamefont
  {Wen}, \citenamefont {Zhu}, \citenamefont {Gong} \emph
  {et~al.}}]{zhu2019generalized}%
  \BibitemOpen
  \bibfield  {author} {\bibinfo {author} {\bibfnamefont {T.}~\bibnamefont
  {Zhu}}, \bibinfo {author} {\bibfnamefont {Y.}~\bibnamefont {Lou}}, \bibinfo
  {author} {\bibfnamefont {Y.}~\bibnamefont {Zhou}}, \bibinfo {author}
  {\bibfnamefont {J.}~\bibnamefont {Zhang}}, \bibinfo {author} {\bibfnamefont
  {J.}~\bibnamefont {Huang}}, \bibinfo {author} {\bibfnamefont
  {Y.}~\bibnamefont {Li}}, \bibinfo {author} {\bibfnamefont {H.}~\bibnamefont
  {Luo}}, \bibinfo {author} {\bibfnamefont {S.}~\bibnamefont {Wen}}, \bibinfo
  {author} {\bibfnamefont {S.}~\bibnamefont {Zhu}}, \bibinfo {author}
  {\bibfnamefont {Q.}~\bibnamefont {Gong}},  \emph {et~al.},\ }\href@noop {}
  {\bibfield  {journal} {\bibinfo  {journal} {Physical Review Applied}\
  }\textbf {\bibinfo {volume} {11}},\ \bibinfo {pages} {034043} (\bibinfo
  {year} {2019})}\BibitemShut {NoStop}%
\bibitem [{\citenamefont {Kwon}\ \emph {et~al.}(2020)\citenamefont {Kwon},
  \citenamefont {Cordaro}, \citenamefont {Sounas}, \citenamefont {Polman},\
  and\ \citenamefont {Alu}}]{kwon2020dual}%
  \BibitemOpen
  \bibfield  {author} {\bibinfo {author} {\bibfnamefont {H.}~\bibnamefont
  {Kwon}}, \bibinfo {author} {\bibfnamefont {A.}~\bibnamefont {Cordaro}},
  \bibinfo {author} {\bibfnamefont {D.}~\bibnamefont {Sounas}}, \bibinfo
  {author} {\bibfnamefont {A.}~\bibnamefont {Polman}}, \ and\ \bibinfo {author}
  {\bibfnamefont {A.}~\bibnamefont {Alu}},\ }\href@noop {} {\bibfield
  {journal} {\bibinfo  {journal} {ACS Photonics}\ } (\bibinfo {year}
  {2020})}\BibitemShut {NoStop}%
\bibitem [{\citenamefont {Lou}\ \emph {et~al.}(2020)\citenamefont {Lou},
  \citenamefont {Fang},\ and\ \citenamefont {Ruan}}]{lou2020optical}%
  \BibitemOpen
  \bibfield  {author} {\bibinfo {author} {\bibfnamefont {Y.}~\bibnamefont
  {Lou}}, \bibinfo {author} {\bibfnamefont {Y.}~\bibnamefont {Fang}}, \ and\
  \bibinfo {author} {\bibfnamefont {Z.}~\bibnamefont {Ruan}},\ }\href@noop {}
  {\bibfield  {journal} {\bibinfo  {journal} {arXiv preprint arXiv:2003.10649}\
  } (\bibinfo {year} {2020})}\BibitemShut {NoStop}%
\bibitem [{\citenamefont {Zangeneh-Nejad}\ and\ \citenamefont
  {Fleury}(2019)}]{zangeneh2019topological}%
  \BibitemOpen
  \bibfield  {author} {\bibinfo {author} {\bibfnamefont {F.}~\bibnamefont
  {Zangeneh-Nejad}}\ and\ \bibinfo {author} {\bibfnamefont {R.}~\bibnamefont
  {Fleury}},\ }\href@noop {} {\bibfield  {journal} {\bibinfo  {journal} {Nature
  communications}\ }\textbf {\bibinfo {volume} {10}},\ \bibinfo {pages} {1}
  (\bibinfo {year} {2019})}\BibitemShut {NoStop}%
\bibitem [{\citenamefont {Guo}\ \emph {et~al.}(2018)\citenamefont {Guo},
  \citenamefont {Xiao}, \citenamefont {Minkov}, \citenamefont {Shi},\ and\
  \citenamefont {Fan}}]{guo2018photonic}%
  \BibitemOpen
  \bibfield  {author} {\bibinfo {author} {\bibfnamefont {C.}~\bibnamefont
  {Guo}}, \bibinfo {author} {\bibfnamefont {M.}~\bibnamefont {Xiao}}, \bibinfo
  {author} {\bibfnamefont {M.}~\bibnamefont {Minkov}}, \bibinfo {author}
  {\bibfnamefont {Y.}~\bibnamefont {Shi}}, \ and\ \bibinfo {author}
  {\bibfnamefont {S.}~\bibnamefont {Fan}},\ }\href@noop {} {\bibfield
  {journal} {\bibinfo  {journal} {Optica}\ }\textbf {\bibinfo {volume} {5}},\
  \bibinfo {pages} {251} (\bibinfo {year} {2018})}\BibitemShut {NoStop}%
\bibitem [{\citenamefont {Zhou}\ \emph {et~al.}(2020)\citenamefont {Zhou},
  \citenamefont {Wu}, \citenamefont {Chen}, \citenamefont {Chen}, \citenamefont
  {Chen},\ and\ \citenamefont {Ma}}]{zhou2020analog}%
  \BibitemOpen
  \bibfield  {author} {\bibinfo {author} {\bibfnamefont {Y.}~\bibnamefont
  {Zhou}}, \bibinfo {author} {\bibfnamefont {W.}~\bibnamefont {Wu}}, \bibinfo
  {author} {\bibfnamefont {R.}~\bibnamefont {Chen}}, \bibinfo {author}
  {\bibfnamefont {W.}~\bibnamefont {Chen}}, \bibinfo {author} {\bibfnamefont
  {R.}~\bibnamefont {Chen}}, \ and\ \bibinfo {author} {\bibfnamefont
  {Y.}~\bibnamefont {Ma}},\ }\href@noop {} {\bibfield  {journal} {\bibinfo
  {journal} {Advanced Optical Materials}\ }\textbf {\bibinfo {volume} {8}},\
  \bibinfo {pages} {1901523} (\bibinfo {year} {2020})}\BibitemShut {NoStop}%
\bibitem [{\citenamefont {Isro}\ \emph {et~al.}(2018)\citenamefont {Isro},
  \citenamefont {Iskandar}, \citenamefont {Kivshar},\ and\ \citenamefont
  {Shadrivov}}]{kivshar}%
  \BibitemOpen
  \bibfield  {author} {\bibinfo {author} {\bibfnamefont {S.~D.}\ \bibnamefont
  {Isro}}, \bibinfo {author} {\bibfnamefont {A.~A.}\ \bibnamefont {Iskandar}},
  \bibinfo {author} {\bibfnamefont {Y.~S.}\ \bibnamefont {Kivshar}}, \ and\
  \bibinfo {author} {\bibfnamefont {I.~V.}\ \bibnamefont {Shadrivov}},\
  }\href@noop {} {\bibfield  {journal} {\bibinfo  {journal} {Optics express}\
  }\textbf {\bibinfo {volume} {26}},\ \bibinfo {pages} {32624} (\bibinfo {year}
  {2018})}\BibitemShut {NoStop}%
\bibitem [{\citenamefont {Davis}\ \emph {et~al.}(2019)\citenamefont {Davis},
  \citenamefont {Eftekhari}, \citenamefont {G{\'o}mez},\ and\ \citenamefont
  {Roberts}}]{davis2019metasurfaces}%
  \BibitemOpen
  \bibfield  {author} {\bibinfo {author} {\bibfnamefont {T.}~\bibnamefont
  {Davis}}, \bibinfo {author} {\bibfnamefont {F.}~\bibnamefont {Eftekhari}},
  \bibinfo {author} {\bibfnamefont {D.}~\bibnamefont {G{\'o}mez}}, \ and\
  \bibinfo {author} {\bibfnamefont {A.}~\bibnamefont {Roberts}},\ }\href@noop
  {} {\bibfield  {journal} {\bibinfo  {journal} {Physical review letters}\
  }\textbf {\bibinfo {volume} {123}},\ \bibinfo {pages} {013901} (\bibinfo
  {year} {2019})}\BibitemShut {NoStop}%
\bibitem [{\citenamefont {Klinkenbusch}(1992)}]{Klinkenbusch}%
  \BibitemOpen
  \bibfield  {author} {\bibinfo {author} {\bibfnamefont {L.}~\bibnamefont
  {Klinkenbusch}},\ }\href@noop {} {\bibfield  {journal} {\bibinfo  {journal}
  {Archiv f{\"u}r Elektrotechnik}\ }\textbf {\bibinfo {volume} {75}},\ \bibinfo
  {pages} {261} (\bibinfo {year} {1992})}\BibitemShut {NoStop}%
\bibitem [{\citenamefont {Zouhdi}\ \emph {et~al.}(2008)\citenamefont {Zouhdi},
  \citenamefont {Sihvola},\ and\ \citenamefont {Vinogradov}}]{Zouhdi}%
  \BibitemOpen
  \bibfield  {author} {\bibinfo {author} {\bibfnamefont {S.}~\bibnamefont
  {Zouhdi}}, \bibinfo {author} {\bibfnamefont {A.}~\bibnamefont {Sihvola}}, \
  and\ \bibinfo {author} {\bibfnamefont {A.~P.}\ \bibnamefont {Vinogradov}},\
  }\href@noop {} {\emph {\bibinfo {title} {Metamaterials and plasmonics:
  fundamentals, modelling, applications}}}\ (\bibinfo  {publisher} {Springer
  Science \& Business Media},\ \bibinfo {year} {2008})\BibitemShut {NoStop}%
\bibitem [{\citenamefont {Papas}(2014)}]{Papas}%
  \BibitemOpen
  \bibfield  {author} {\bibinfo {author} {\bibfnamefont {C.~H.}\ \bibnamefont
  {Papas}},\ }\href@noop {} {\emph {\bibinfo {title} {Theory of electromagnetic
  wave propagation}}}\ (\bibinfo  {publisher} {Courier Corporation},\ \bibinfo
  {year} {2014})\BibitemShut {NoStop}%
\bibitem [{\citenamefont {Asadchy}\ \emph {et~al.}(2014)\citenamefont
  {Asadchy}, \citenamefont {Faniayeu}, \citenamefont {Ra’di},\ and\
  \citenamefont {Tretyakov}}]{Asadchy}%
  \BibitemOpen
  \bibfield  {author} {\bibinfo {author} {\bibfnamefont {V.~S.}\ \bibnamefont
  {Asadchy}}, \bibinfo {author} {\bibfnamefont {I.~A.}\ \bibnamefont
  {Faniayeu}}, \bibinfo {author} {\bibfnamefont {Y.}~\bibnamefont {Ra’di}}, \
  and\ \bibinfo {author} {\bibfnamefont {S.~A.}\ \bibnamefont {Tretyakov}},\
  }\href@noop {} {\bibfield  {journal} {\bibinfo  {journal} {Photonics and
  Nanostructures-Fundamentals and Applications}\ }\textbf {\bibinfo {volume}
  {12}},\ \bibinfo {pages} {298} (\bibinfo {year} {2014})}\BibitemShut
  {NoStop}%
\bibitem [{\citenamefont {Yazdi}\ and\ \citenamefont
  {Komjani}(2016{\natexlab{a}})}]{Yazdi1}%
  \BibitemOpen
  \bibfield  {author} {\bibinfo {author} {\bibfnamefont {M.}~\bibnamefont
  {Yazdi}}\ and\ \bibinfo {author} {\bibfnamefont {N.}~\bibnamefont
  {Komjani}},\ }\href@noop {} {\bibfield  {journal} {\bibinfo  {journal} {JOSA
  B}\ }\textbf {\bibinfo {volume} {33}},\ \bibinfo {pages} {491} (\bibinfo
  {year} {2016}{\natexlab{a}})}\BibitemShut {NoStop}%
\bibitem [{\citenamefont {Mirmoosa}\ \emph {et~al.}(2014)\citenamefont
  {Mirmoosa}, \citenamefont {Ra’di}, \citenamefont {Asadchy}, \citenamefont
  {Simovski},\ and\ \citenamefont {Tretyakov}}]{Mirmoosa}%
  \BibitemOpen
  \bibfield  {author} {\bibinfo {author} {\bibfnamefont {M.}~\bibnamefont
  {Mirmoosa}}, \bibinfo {author} {\bibfnamefont {Y.}~\bibnamefont {Ra’di}},
  \bibinfo {author} {\bibfnamefont {V.}~\bibnamefont {Asadchy}}, \bibinfo
  {author} {\bibfnamefont {C.}~\bibnamefont {Simovski}}, \ and\ \bibinfo
  {author} {\bibfnamefont {S.}~\bibnamefont {Tretyakov}},\ }\href@noop {}
  {\bibfield  {journal} {\bibinfo  {journal} {Physical Review Applied}\
  }\textbf {\bibinfo {volume} {1}},\ \bibinfo {pages} {034005} (\bibinfo {year}
  {2014})}\BibitemShut {NoStop}%
\bibitem [{\citenamefont {Yazdi}\ and\ \citenamefont
  {Komjani}(2016{\natexlab{b}})}]{Yazdi}%
  \BibitemOpen
  \bibfield  {author} {\bibinfo {author} {\bibfnamefont {M.}~\bibnamefont
  {Yazdi}}\ and\ \bibinfo {author} {\bibfnamefont {N.}~\bibnamefont
  {Komjani}},\ }\href@noop {} {\bibfield  {journal} {\bibinfo  {journal}
  {Progress In Electromagnetics Research}\ }\textbf {\bibinfo {volume} {45}},\
  \bibinfo {pages} {123} (\bibinfo {year} {2016}{\natexlab{b}})}\BibitemShut
  {NoStop}%
\bibitem [{\citenamefont {Dasgupta}\ and\ \citenamefont {Fuchs}(1981)}]{Basab}%
  \BibitemOpen
  \bibfield  {author} {\bibinfo {author} {\bibfnamefont {B.~B.}\ \bibnamefont
  {Dasgupta}}\ and\ \bibinfo {author} {\bibfnamefont {R.}~\bibnamefont
  {Fuchs}},\ }\href@noop {} {\bibfield  {journal} {\bibinfo  {journal}
  {Physical Review B}\ }\textbf {\bibinfo {volume} {24}},\ \bibinfo {pages}
  {554} (\bibinfo {year} {1981})}\BibitemShut {NoStop}%
\bibitem [{\citenamefont {Safari}\ \emph {et~al.}(2018)\citenamefont {Safari},
  \citenamefont {Albooyeh}, \citenamefont {Simovski},\ and\ \citenamefont
  {Tretyakov}}]{safari}%
  \BibitemOpen
  \bibfield  {author} {\bibinfo {author} {\bibfnamefont {M.}~\bibnamefont
  {Safari}}, \bibinfo {author} {\bibfnamefont {M.}~\bibnamefont {Albooyeh}},
  \bibinfo {author} {\bibfnamefont {C.}~\bibnamefont {Simovski}}, \ and\
  \bibinfo {author} {\bibfnamefont {S.}~\bibnamefont {Tretyakov}},\ }\href@noop
  {} {\bibfield  {journal} {\bibinfo  {journal} {Physical Review B}\ }\textbf
  {\bibinfo {volume} {97}},\ \bibinfo {pages} {085412} (\bibinfo {year}
  {2018})}\BibitemShut {NoStop}%
\bibitem [{\citenamefont {Alu}\ and\ \citenamefont {Engheta}(2010)}]{Alu}%
  \BibitemOpen
  \bibfield  {author} {\bibinfo {author} {\bibfnamefont {A.}~\bibnamefont
  {Alu}}\ and\ \bibinfo {author} {\bibfnamefont {N.}~\bibnamefont {Engheta}},\
  }\href@noop {} {\bibfield  {journal} {\bibinfo  {journal} {Journal of
  Nanophotonics}\ }\textbf {\bibinfo {volume} {4}},\ \bibinfo {pages} {041590}
  (\bibinfo {year} {2010})}\BibitemShut {NoStop}%
\bibitem [{\citenamefont {Palik}(1998)}]{palik1998handbook}%
  \BibitemOpen
  \bibfield  {author} {\bibinfo {author} {\bibfnamefont {E.~D.}\ \bibnamefont
  {Palik}},\ }\href@noop {} {\emph {\bibinfo {title} {Handbook of optical
  constants of solids}}},\ Vol.~\bibinfo {volume} {3}\ (\bibinfo  {publisher}
  {Academic press},\ \bibinfo {year} {1998})\BibitemShut {NoStop}%
\bibitem [{\citenamefont {Albooyeh}\ \emph {et~al.}(2017)\citenamefont
  {Albooyeh}, \citenamefont {Kwon}, \citenamefont {Capolino},\ and\
  \citenamefont {Tretyakov}}]{albooyeh2017equivalent}%
  \BibitemOpen
  \bibfield  {author} {\bibinfo {author} {\bibfnamefont {M.}~\bibnamefont
  {Albooyeh}}, \bibinfo {author} {\bibfnamefont {D.-H.}\ \bibnamefont {Kwon}},
  \bibinfo {author} {\bibfnamefont {F.}~\bibnamefont {Capolino}}, \ and\
  \bibinfo {author} {\bibfnamefont {S.}~\bibnamefont {Tretyakov}},\ }\href@noop
  {} {\bibfield  {journal} {\bibinfo  {journal} {Physical Review B}\ }\textbf
  {\bibinfo {volume} {95}},\ \bibinfo {pages} {115435} (\bibinfo {year}
  {2017})}\BibitemShut {NoStop}%
\bibitem [{\citenamefont {Laroche}\ \emph
  {et~al.}(2006{\natexlab{b}})\citenamefont {Laroche}, \citenamefont
  {Albaladejo}, \citenamefont {G{\'o}mez-Medina},\ and\ \citenamefont
  {S{\'a}enz}}]{laroche2006tuning}%
  \BibitemOpen
  \bibfield  {author} {\bibinfo {author} {\bibfnamefont {M.}~\bibnamefont
  {Laroche}}, \bibinfo {author} {\bibfnamefont {S.}~\bibnamefont {Albaladejo}},
  \bibinfo {author} {\bibfnamefont {R.}~\bibnamefont {G{\'o}mez-Medina}}, \
  and\ \bibinfo {author} {\bibfnamefont {J.~J.}\ \bibnamefont {S{\'a}enz}},\
  }\href@noop {} {\bibfield  {journal} {\bibinfo  {journal} {Physical Review
  B}\ }\textbf {\bibinfo {volume} {74}},\ \bibinfo {pages} {245422} (\bibinfo
  {year} {2006}{\natexlab{b}})}\BibitemShut {NoStop}%
\bibitem [{\citenamefont {Niemi}\ \emph {et~al.}(2013)\citenamefont {Niemi},
  \citenamefont {Karilainen},\ and\ \citenamefont
  {Tretyakov}}]{niemi2013synthesis}%
  \BibitemOpen
  \bibfield  {author} {\bibinfo {author} {\bibfnamefont {T.}~\bibnamefont
  {Niemi}}, \bibinfo {author} {\bibfnamefont {A.~O.}\ \bibnamefont
  {Karilainen}}, \ and\ \bibinfo {author} {\bibfnamefont {S.~A.}\ \bibnamefont
  {Tretyakov}},\ }\href@noop {} {\bibfield  {journal} {\bibinfo  {journal}
  {IEEE Transactions on Antennas and Propagation}\ }\textbf {\bibinfo {volume}
  {61}},\ \bibinfo {pages} {3102} (\bibinfo {year} {2013})}\BibitemShut
  {NoStop}%
\end{thebibliography}%






\end{document}